\documentclass[12pt,compress]{elsarticle}
\usepackage{amsmath,amssymb,amscd}
\usepackage{color}
\journal{Nuclear Physics B}

\newcommand{\beq}{\begin{equation}}
\newcommand{\eeq}{\end{equation}}
\newcommand{\bea}{\begin{eqnarray}}
\newcommand{\eea}{\end{eqnarray}}
\newcommand{\bse}{\begin{subequations}}
\newcommand{\ese}{\end{subequations}}

\newcommand{\noi}{\noindent}

\newcommand{\nn}{\nonumber}

\newcommand{\ba}{\begin{array}}
\newcommand{\ea}{\end{array}}
\newcommand{\balign}{\begin{align}}
\newcommand{\ealign}{\end{align}}

\newcommand{\be}{\beta}
\newcommand{\de}{\delta}
\newcommand{\ep}{\epsilon}

\newcommand{\mbf}[1]{\mathbf{#1}}

\newcommand{\mbs}[1]{\boldsymbol{#1}}

\newcommand{\mbb}[1]{\mathbb{#1}}

\newcommand{\mc}[1]{\mathcal{#1}}

\newcommand{\mr}[1]{\mathrm{#1}}

\newcommand{\wt}[1]{\widetilde{#1}}

\newcommand{\one}{1\hskip -1mm{\rm l}}

\renewcommand{\d}{\partial}
\renewcommand{\le}{\leqslant}

\renewcommand{\leq}{\leqslant}
\renewcommand{\geq}{\geqslant}

\renewcommand{\v}{\vert}
\newcommand{\half}{$\frac{1}{2}~$}
\newcommand{\su}[1]{su($#1$)}

\newcommand\ket[1]{|#1\rangle}

\newcommand{\qbinom}[3]{\genfrac{[}{]}{0pt}{}{\,#1\,}{#2}_{#3}}

\newcommand{\eq}[1]{(\ref{#1})}
\newcommand{\Eq}[1]{Eq.~(\ref{#1})}
\newcommand{\tr}{\operatorname{tr}}

\begin{document}

\begin{center}
{\Large \bf \sf 
Supersymmetric analogue of {$BC_N$} type rational integrable
models with polarized spin reversal operators}

\vspace{1.3cm}

{\sf 
P. Banerjee$^1$\footnote{E-mail address: pratyay.banerjee@saha.ac.in},
B. Basu-Mallick$^1$\footnote{
Corresponding author, Fax: +91-33-2337-4637, 
Telephone: +91-33-2337-5345, 
E-mail address: bireswar.basumallick@saha.ac.in},
N. Bondyopadhaya$^2$\footnote{E-mail address:
nilanjan.iserc@visva-bharati.ac.in; 
Permanent address: Integrated Science Education and Research Centre,
Visva-Bharati University, Santiniketan 731 235, India}
and C. Datta$^1$
\footnote{E-mail address: chitralekha.datta@saha.ac.in}}
\bigskip

{\em $^1$Theory Division, Saha Institute of Nuclear Physics, \\
1/AF Bidhan Nagar, Kolkata 700 064, India}

\bigskip
{\em $^2$BLTP, Joint Institute of Nuclear Research, 
Dubna, Moscow region, 141980, Russia}


\end{center}

\bigskip
\bigskip

\noi {\bf Abstract}
\vspace {.2 cm}

We derive the exact spectra as well as partition functions for a class  
of $BC_N$ type of spin Calogero models, whose Hamiltonians
are constructed by using supersymmetric
analogues of polarized spin reversal operators (SAPSRO).
The strong coupling limit of these spin Calogero models yields   
$BC_N$ type of Polychronakos-Frahm (PF) spin chains with SAPSRO.
By applying the freezing trick, we obtain an exact expression
for the partition functions of such PF spin 
chains. We also derive a formula which 
expresses the partition function of any $BC_N$ type of PF spin chain  
with SAPSRO in terms of partition functions of several $A_{K}$
type of supersymmetric PF spin chains, where $K\leq N-1$. 
Subsequently we show that an extended boson-fermion duality relation
is obeyed by the partition functions of 
the $BC_N$ type of PF chains with SAPSRO.  
Some spectral properties of these spin chains, 
like level density distribution and    
nearest neighbour spacing distribution, are also studied.

\vspace{.74 cm}
\noi PACS No. : 02.30.Ik, 05.30.-d, 75.10.Pq, 75.10.Jm 

\vspace {.2 cm}

\newpage

\baselineskip 16 pt

\noi \section{Introduction}
\renewcommand{\theequation}{1.{\arabic{equation}}}
\setcounter{equation}{0}
\medskip
Remarkable progress has been made in recent years in the  
computation of exact spectra, partition functions and correlation functions
of one-dimensional quantum integrable spin chains with long-range 
interactions as well as their supersymmetric 
generalizations~\cite{Ha88,Sh88,HHTBP92,BGHP93,Ha93, 
Po93,Fr93,Po94,Hi95npb,BUW99,
HB00,FG05,BB06,BBS08,BFGR08epl,EFG10, 
BBH10,SA94,BPS95,YT96,EFGR05,BFGR08,Ha96,KK09}.  
Exact solutions of this type of quantum spin chains 
with periodic and open boundary conditions have been found to be 
closely connected with diverse areas of physics and mathematics 
like condensed matter systems exhibiting generalized exclusion
statistics~\cite{Ha93,Ha96,KK09,Po06}, quantum Hall effect~\cite{AI94}, 
quantum electric transport phenomena~\cite{BR94,Ca95},
calculation of higher loop effects in the spectra of trace operators 
of planar ${\mathcal N}=4$ super Yang--Mills theory~\cite{BKS03,Be04,BBL09},
Dunkl operators related to various root systems ~\cite{Du98,FGGRZ01}, 
random matrix theory~\cite{TSA95},
and Yangian quantum groups~\cite{BGHP93,Ha93,Hi95npb,BBH10,Ba99,BBHS07,BE08}.
Furthermore, it has been recently observed that exactly solvable spin 
chains with long-range interactions can be generated through  
some lattice discretizations of conformal field 
theories related to the `infinite  
matrix product states'~\cite{CS10,NCS11,TNS14,BQ14}. 
 
The study of quantum integrable spin chains with long-range interactions 
was pioneered by Haldane and Shastry, who derived the 
exact spectrum of a spin-\half  
chain with lattice sites equally spaced on a circle and 
spins interacting 
through pairwise exchange interactions inversely proportional 
to the square of their chord distances~\cite{Ha88,Sh88}. 
It has been found that, 
the exact ground state wave function
of this \su{2} symmetric Haldane-Shastry (HS) spin
chain  coincides with 
the $U\to\infty$ limit of Gutzwiller's variational wave function 
describing the ground state of the
one-dimensional Hubbard model~\cite{Gu63,GV87,GJR87}. 
A close relation between the 
\su{m} generalizations of this HS spin chain and the 
(trigonometric) Sutherland model  
has been established by using the 
`freezing trick'~\cite{Po93,SS93},  
which we briefly describe in the following.
In contrast to the case of HS spin chain where lattice sites are  
fixed at equidistant positions on a circle, 
the particles of the \su{m} spin Sutherland model can move on a circle 
and they contain both coordinate as 
well as spin degrees of freedom. However, 
in the strong coupling limit, the coordinates of these particles
decouple from their spins and 
`freeze' at the minimum value of the scalar part of the 
potential. Furthermore, this minimum value of the scalar part of the 
potential yields the equally spaced
lattice points of the HS spin chain. As a result, 
in the strong coupling limit, 
the dynamics of the decoupled spin degrees of freedom of 
the \su{m} spin Sutherland model is governed by
the Hamiltonian of the \su{m} HS model. 
Application of this freezing trick to the \su{m} spin
(rational) Calogero model 
leads to another quantum integrable spin chain with
long-range interaction~\cite{Po93}, which is 
known in the literature as the \su{m}
Polychronakos or Polychronakos--Frahm (PF) spin chain.
The sites of such rational PF spin chain  
are inhomogeneously spaced on a line and, in fact, they coincide    
with the zeros of the Hermite polynomial~\cite{Fr93}.
Indeed, the Hamiltonian of the \su{m}  PF spin chain 
is given by    
\beq
\label{a1}
\mc{H}_\mr{PF}^{(m)}=\sum_{1\leq i< j \leq N}
\frac{1- \ep P_{ij}^{(m)}}{(\rho_i-\rho_j)^2}  \, ,
\eeq
where 
$\ep = 1 ~(-1)$ corresponds to the 
ferromagnetic (anti-ferromagnetic) case, 
$P_{ij}^{(m)}$ denotes the exchange operator 
which interchanges the `spins' 
(taking $m$ possible values) of $i$-th and $j$-th lattice sites
and $\rho_i$ denotes the $i$-th zero of the 
Hermite polynomial of degree $N$.
Due to the decoupling of  
the spin and coordinate degrees of freedom of the \su{m} spin Calogero model
for large values of its coupling constant, 
an exact expression for    
the partition function of \su{m} PF spin chain 
can be derived by dividing the partition function of the 
\su{m} spin Calogero model through  
that of the spinless Calogero model~\cite{Po94}.
Similarly, the partition function of \su{m} HS spin chain 
can be computed by dividing the partition function of the 
\su{m} spin Sutherland model through  
that of the spinless Sutherland model~\cite{FG05}.

As is well known, supersymmetric spin chains with different 
type of interactions
play an important role in describing some quantum 
impurity problems and disordered systems 
in condensed matter physics, 
where holes moving in the dynamical background of spins behave as bosons, 
and spin-1/2 electrons behave as fermions \cite{Sc97,Sa00,EFS05,ASK01,TSG07}.
The above mentioned PF and HS spin chains admit natural 
\su{m\v n} supersymmetric generalizations, where each lattice site
has $m$ number of bosonic and $n$ number of fermionic degrees of freedom. 
Exact expressions for the partition functions 
of such \su{m\v n} PF and HS spin chains can also be computed 
by using the method of freezing trick \cite{BUW99,HB00,BB06}. 
It is found that  
these partition functions satisfy remarkable duality relations 
under the exchange of bosonic and fermionic degrees of freedom. 

It may be noted that, the strength of interaction between any two 
spins in the Hamiltonian \eq{a1} 
depends only on the difference of their site coordinates.
This type of translationally invariant Hamiltonians of 
quantum integrable spin chains 
(and their supersymmetric generalizations)
are closely related to the $A_{N-1}$ type of root system.
Indeed, the spin-spin interactions appearing in such Hamiltonians 
are given by the permutation operators which yield a realization of the 
$A_{N-1}$ type of Weyl group. 
However, it is also possible to construct exactly solvable 
variants of HS and PF spin chains associated with the 
$BC_N$, $B_N$, $C_N$ and $D_N$ 
root systems~\cite{SA94,BPS95,YT96,EFGR05,BFGR08,BFG09,
BFG11,BFG13}.
A key feature of such spin chains is the presence of boundary points
with reflecting mirrors, due to which the spins not only interact
with each other but also with their mirror images. 
As a result, the corresponding  
Hamiltonians break the translational invariance.
It may also be noted that, Hamiltonians of the spin chains  
associated with the $BC_N$ root system and 
its $B_N$, $C_N$ and $D_N$ degenerations contain 
reflection operators like $S_i$ ($i=1,\dots,N$),
which satisfy the relation $S_i^2=\one$ and few other relations 
associated with the corresponding Weyl algebra.  
Representing such $S_i$ as the spin reversal operator 
$P_i$ which changes the sign of 
the spin component on the $i$-th lattice site,  
the partition functions of 
HS and PF spin chains associated with the 
$BC_N$, $B_N$, $C_N$ and $D_N$ 
root systems have been computed by
using the freezing trick~\cite{EFGR05,BFGR08,BFG09,
BFG11,BFG13}. Furthermore, by  
taking $S_i$ as the spin reversal operator on a superspace,
the partition function of 
a supersymmetric analogue of the PF spin chain associated with  
$BC_N$ root system has also been computed in a similar way~\cite{BFGR09}.

However it is worth noting that, 
the above mentioned representations of reflection 
operators as the spin reversal 
operators is by no means the only possible choice. 
Indeed, by choosing all   
reflection operators as the trivial identity operator,  
it has been found that~\cite{BPS95} a spin-\half 
HS chain associated with 
the $BC_N$ root system leads to an integrable 
\su{2} invariant spin model   
which was first studied by Simons and Altshuler \cite{SA94}.
Furthermore, a class of exactly solvable 
spin Calogero models of $BC_N$ type and the corresponding 
PF chains have been introduced recently~\cite{BBB14}, 
where the reflection operators are represented by 
arbitrarily polarized spin
reversal operators (PSRO) $P_i^{(m_1,m_2)}$, 
which act as the identity on the first $m_1$ elements
of the spin basis on the $i$-th lattice site and 
as minus the identity on the rest of the spin basis. 
Consequently, depending on the action of $P_i^{(m_1,m_2)}$, 
the basis vectors of the $m$-dimensional spin space on each 
lattice site can be grouped into two cases --- 
$m_1$ elements with 
positive parity and $m_2$ elements with negative parity.
Using a similarity transformation, it can be shown that the PSRO  
reduce to the usual spin reversal operators 
$P_i$ (up to a sign factor) when $m_1=m_2$ or
$m_1=m_2\pm1$.
For the remaining values of the discrete
parameters $m_1$ and $m_2$, the systems constructed 
in the later reference differ from the
standard Calogero and PF models of $BC_N$-type. 
In particular, for the case $m_2=0$ and $m_1= m$,
$P_i$ reduces to the identity operator and 
leads to a novel 
\su{m} invariant spin chain, 
which is described by the Hamiltonian 
\beq
\label{a2}
{\mc{H}}^{(m,0)}= \sum_{1\leq i \neq j \leq N}
 \frac{y_i+y_j}{(y_i-y_j)^2} \, 
  (1- \ep P_{ij}^{(m)})  \, ,
\eeq
where 
$\ep = \pm 1$, 
$y_i$ denotes the $i$-th zero  
of the generalized Laguerre polynomial $L_N^{\beta -1}$.
Thus, the lattice sites of ${\mc{H}}^{(m,0)}$
implicitly depend on the real positive parameter $\beta$.
Computing the partition function of the spin chain \eq{a2}
by using the freezing trick and analyzing 
such partition function, it has been found that    
the spectrum of this spin chain coincides (up to a scale factor)
with that of the original PF model \eq{a1} \cite{BBB14}.
However, a deeper reason for this surprising coincidence 
has not been fully understood till now. 

Even though the spectrum and partition function of the 
supersymmetric generalization of the 
$A_{N-1}$ type of PF spin chain \eq{a1}
have been computed earlier~\cite{BUW99,HB00},
no such result is available till now 
for the supersymmetric generalization of the spin chain \eq{a2}.
In this context it is interesting to ask whether it is possible
to compute the partition function for the
supersymmetric version of the spin chain \eq{a2} by using 
the freezing trick, and whether the corresponding spectrum  
can be related in a simple way with  
that of the supersymmetric PF spin chain.  
In the present article we try to answer these questions 
by constructing supersymmetric analogues of  
PSRO (SAPSRO), which would satisfy the $BC_N$ type of Weyl algebra.
By using such SAPSRO,  
we obtain a rather large class of exactly solvable  
spin Calogero models and PF chains of $BC_N$ type. 
In a particular case where polarization is minimal, 
SAPSRO reduce to the supersymmetric analogues of usual 
spin reversal operators 
and lead to the spin Calogero models as well as PF chains 
of $BC_N$ type which have been studied earlier~\cite{BFGR09}. However, 
in all other cases, these SAPSRO can be used 
to generate novel exactly solvable 
spin Calogero models and PF chains of $BC_N$ type. In particular, 
for the case where polarization is maximal, we find that  
SAPSRO reduces to the trivial
identity operator and lead to a supersymmetric extension 
of the spin chain \eq{a2}, whose partition function and spectrum 
can be computed by using the freezing trick. 

Another interesting topic which we shall address in this paper
is a modification of the usual boson-fermion duality relation
which is satisfied by the partition functions of $A_{N-1}$ type 
of spin chains. This type of modified duality relation
has been studied earlier for the special case of $BC_N$ type 
of PF chains 
associated with the supersymmetric analogue
of the spin reversal operators~\cite{BFGR09}. 
It has been observed that this duality relation not only  
involves the exchange of bosonic and fermionic 
degrees freedom, but also certain changes of the two discrete parameters
which appear in the corresponding Hamiltonian.
However, the full significance for such change 
of the two discrete parameters 
has not been explored till now. We find that the underlying reason 
for such change of the discrete parameters can be understood   
in a natural way if one studies 
the duality relation for $BC_N$ type of PF chains
in the broader context of SAPSRO. 
Indeed, in this paper we consider a new quantum number
which measures the parity of the spin states under the action of SAPSRO.
Curiously, it turns out that the partition functions of the spin chains
now satisfy an `extended' boson-fermion duality relation,  
which involves not only the exchange of bosonic and 
fermionic degrees of freedom, but also the exchange of           
positive and negative parity degrees of freedom associated with 
the SAPSRO. 

The arrangement of this paper is as follows. In Section~2,
we construct SAPSRO which, 
along with the supersymmetric spin exchange operators, lead to     
new representations of the $BC_N$ type of Weyl algebra
and related PF spin chains with 
open boundary conditions.  Next, in Section~3, we consider   
$BC_N$ type of spin Calogero models associated with 
SAPSRO, which in the strong coupling limit yield the above 
mentioned class of PF spin chains. We derive the exact 
spectra as well as partition functions of these 
$BC_N$ type of spin Calogero models with SAPSRO. 
By applying the freezing trick, subsequently
we obtain an exact expression
for the partition functions of the related PF spin 
chains. In Section~4, we derive a formula which 
expresses the partition function of any $BC_N$ type of PF spin chain  
with SAPSRO in terms of partition functions of several $A_{K}$
type of supersymmetric PF spin chains, where $K\leq N-1$. 
By taking a particular 
limit of the above mentioned formula, we find that the partition
function of the supersymmetric extension 
of the spin chain \eq{a2} coincides with that of a $A_{N-1}$
type of supersymmetric PF spin chain. In Section~5, we derive 
an extended boson-fermion duality relation for the $BC_N$ type of
PF chains with SAPSRO.  
In Section~6, we compute the ground state and the highest 
state energies of these spin 
chains. Some spectral properties of these spin chains, 
like level density distribution and    
nearest neighbour spacing distribution, are studied in Section~7. 
Section~8 is the concluding section.

\noi \section{$BC_N$ type of Weyl algebra and related PF chains}
\renewcommand{\theequation}{2.{\arabic{equation}}}
\setcounter{equation}{0}
\medskip
As is well known, different representations of the  
$BC_N$ type of Weyl algebra play a key role in constructing
exactly solvable variants of 
HS and PF spin chains with open boundary conditions. 
This $BC_N$ type of Weyl algebra is 
generated by the elements $\mc{W}_{ij}$ and $\mc{W}_{i}\,$ 
satisfying the relations 
\begin{subequations}
\bea
&\mc{W}_{ij}^2=\one \, ,
~~~~\mc{W}_{ij}\mc{W}_{jk} =\mc{W}_{ik}\mc{W}_{ij}
=\mc{W}_{jk}\mc{W}_{ik} \, ,~~~~ \mc{W}_{ij}\mc{W}_{kl}
=\mc{W}_{kl}\mc{W}_{ij} \, ,  \label{weyl1}\\ 
&\mc{W}_{i}^2=\one \, , 
~~~~ \mc{W}_{i}\mc{W}_{j}= \mc{W}_{j}\mc{W}_{i} \, ,
~~~~ \mc{W}_{ij}\mc{W}_{k}=\mc{W}_{k}\mc{W}_{ij} \, , 
~~~~\mc{W}_{ij}\mc{W}_{j}=\mc{W}_{i}\mc{W}_{ij} \, , \label{weyl2}
\eea
\label{b1}
\end{subequations}
where $i,~j,~k,~l$ are all different indices. 
Let us assume that
the Hermitian operators $\mc{P}_{ij}$ and $\mc{P}_{i}\,$
yield a realization 
of the elements $\mc{W}_{ij}$ and $\mc{W}_{i}\,$ respectively  
on an appropriate spin space. 
Motivated by the earlier works~\cite{YT96,BFGR08,BFGR09,BBB14}, 
we define a general form of Hamiltonian for the  
$BC_N$ type of PF spin chain as 
\beq
\mathcal{H}
=\sum_{i\neq j}\left[\frac{1- \mc{P}_{ij}}{(\xi_i-\xi_j)^2} +
\frac{1-\widetilde{\mc{P}}_{ij}}{(\xi_i+\xi_j)^2}\right]
+\beta\sum_{i=1}^{N}\frac{1-\mc{P}_i}{\xi_i^2} \, ,  
\label{b1a}
\eeq
where 
$\beta$ is a positive parameter, 
$ \widetilde{\mc{P}}_{ij}= \mc{P}_i\mc{P}_j \mc{P}_{ij}$, 
$\xi_i=\sqrt{2y_i}$ and $y_i$ 
represents the $i$-th zero point  
of the generalized Laguerre polynomial $L_N^{\beta -1}$.
In the following, at first we shall briefly discuss how this general 
form of Hamiltonian yields already known  
PF spin chains associated with the $BC_N$ root system
for different choices of 
the operators $\mc{P}_{ij}$ and $\mc{P}_{i}$. 
Subsequently, we shall construct SAPSRO which, 
along with the supersymmetric spin exchange operators, would lead to
a new class of representations for the $BC_N$ type of Weyl algebra
and the related PF chains.

In the case of a non-supersymmetric spin chain
with $N$ number of lattice sites, 
the total internal space $\mbs{\Sigma}^{(m)}$ is expressed as 
\beq
\mbs{\Sigma}^{(m)}
\equiv \underbrace{\mc{C}_m \otimes \mc{C}_m \otimes \cdots 
\otimes \mc{C}_m}_{N} \, ,
\label{b2}
\eeq
where $\mc{C}_m$ denotes a $m$-dimensional complex vector space. 
In terms of orthonormal basis vectors, 
$\mbs{\Sigma}^{(m)}$ may be written as     
\beq
\mbs{\Sigma}^{(m)}
= \Big{\langle} \ket{s_1,\cdots,s_N} ~  \Big{\vert} 
s_i\in \{-M,-M+1, \cdots , M \}; ~ M= \frac{m-1}{2} \Big{\rangle} \, . 
\label{b3}
\eeq
The spin exchange operator $P_{ij}^{(m)}$ 
and the spin reversal operator 
$P_i$ act on these orthonormal basis vectors as  
\begin{subequations}
\bea
&&P_{ij}^{(m)}\ket{s_1 \, , \cdots, s_i \, , 
\cdots, s_j \, , \cdots, s_N}
= 
\ket{s_1 \, , \cdots, s_j \, , \cdots, s_i \, , \cdots, s_N} \, , 
\label{spin1} \\
&&P_{i}\ket{s_1 \, , \cdots, s_i \, , \cdots, s_N}
= \ket{s_1 \, , \cdots, - s_i \, , \cdots, s_N} \, .
\label{b4}
\eea
\label{b5}
\end{subequations}
It is easy to check that $\ep P_{ij}^{(m)}$ and $\ep' P_i$ 
(where $\ep,\ep^{\prime}=\pm 1$ are two independent signs)
yield a realization 
of the  $BC_N$ type of Weyl algebra \eq{b1}. 
Substituting $\ep P_{ij}^{(m)}$ and $\ep' P_i$ 
in the places of $\mc{P}_{ij}$ and $\mc{P}_{i}$ respectively in 
the general form of Hamiltonian \eq{b1a}, one obtains an exactly 
solvable $BC_N$ type of non-supersymmetric PF spin chain 
whose partition function has been computed  
by using the freezing trick~\cite{BFGR08}. 

For the purpose of generalizing the above mentioned spin chain 
through PSRO, it is convenient to define the space 
$\mbs{\Sigma}^{(m)}$ through 
a different set of orthonormal basis vectors as
\beq
\mbs{\Sigma}^{(m)}
= \Big{\langle} \ket{s_1,\cdots,s_N} ~  \Big{\vert} 
s_i \in \{ 1,2, \cdots , m \} \Big{\rangle} \, .  
\label{b5a}
\eeq
The action of spin exchange operator $P_{ij}^{(m)}$  
on these orthonormal basis vectors 
is again given by an equation of the form \eq{spin1}.  However, the 
spin reversal operator is replaced by PSRO 
(denoted by $P_i^{(m_1,m_2)}$ for the $i$-th lattice site) 
which acts on these orthonormal basis vectors as~\cite{BBB14}
\beq
P_i^{(m_1,m_2)}\ket{s_1,\cdots,s_i,\cdots,s_N}
=(-1)^{f(s_i)}\ket{s_1,\cdots,s_i,\cdots,s_N},
\label{b6}
\eeq
where
\begin{equation}
 f(s_i) = \left \{  
\begin{array}{ll} 
0,  & \mbox{~if  $s_i\in \{1,2,\cdots,m_1\}$, } \\
1, &   \mbox{~if $s_i\in \{m_1+1,\cdots,m_1+m_2\}$, } \nonumber
\end{array} 
\right.
\label{b7}
\end{equation}
and $m_1$ and $m_2$ are two arbitrary non-negative integers 
satisfying the relation $m_1+m_2=m$. 
Using Eqs.~\eq{spin1} and \eq{b6}, it is easy to check that 
$\ep P_{ij}^{(m)}$ and $P_i^{(m_1,m_2)}$ yield a realization of $BC_N$
type of Weyl algebra \eq{b1}.
Substituting $\ep P_{ij}^{(m)}$  and $P_i^{(m_1,m_2)}$ 
(in places of $\mc{P}_{ij}$ and $\mc{P}_{i}$, respectively) 
in the general form of Hamiltonian \eq{b1a} and taking 
different possible values of $m_1$ and $m_2$, one obtains 
a class of exactly solvable $BC_N$ type of PF spin chains 
with PSRO~\cite{BBB14}. Using a similarity transform it has been shown 
in the latter reference that,  in the special case given by 
$m_1=m_2$ ($m_1=m_2+\ep'$) for even (odd) values of $m$, 
the operator $P_i^{(m_1,m_2)}$ becomes equivalent
to $\ep'{P}_{i}$.
Consequently, PF spin chain associated 
with PSRO reduces to PF spin chain associated 
with spin reversal operators in this special case. 
It may also be observed that, in another
special case given by $m_1=m,~m_2=0$, 
$P_i^{(m_1,m_2)}$ in \eq{b6} reduces to the trivial identity operator
and the corresponding Hamiltonian \eq{b1a}
yields the exactly solvable $su(m)$ invariant spin chain 
\eq{a2} which has been discussed earlier. 

Next, for the purpose of discussing 
representations of the $BC_N$ type of Weyl 
algebra \eq{b1} on a superspace,  
we consider a set of operators like
$C_{j \alpha}^\dagger$($C_{j \alpha}$) which creates (annihilates)
a particle of species $\alpha$ on the $j$-th lattice site.
The parity of these operators are defined as 
\bea
 &&\pi(C_{j \alpha})=\pi(C_{j \alpha}^\dagger)=0 ~
\mr{for}~ \alpha \in [1,2,....,m] \, , \nn \\
 &&\pi(C_{j \alpha})=\pi(C_{j \alpha}^\dagger)=1 ~
 \mr{for}~ \alpha \in [m+1,m+2,....,m+n] \, , \nn
\eea
i.e, they are assumed to be bosonic when 
$\alpha \in [1,2,....,m]$ and 
fermionic when $\alpha \in [m+1,m+2,....,m+n]$.
These operators satisfy commutation (anti-commutation) relations 
given by 
\beq
[C_{j \alpha},C_{k \beta}]_{\pm}=0 \, ,~ 
[C_{j \alpha}^\dagger,C_{k \beta}^\dagger]_{\pm}=0 \, , ~
[C_{j \alpha},C_{k \beta}^\dagger]_{\pm}=\delta_{jk}\delta_{\alpha \beta} \, ,
\label{b7a}
\eeq
where $[C,D]_{\pm} \equiv CD- (-1)^{\pi(C)\pi(D)}DC$.
On a subspace of the corresponding Fock space,  
where each lattice site is occupied by only one particle (i.e., 
$\sum_{\alpha=1}^{m+n} C_{j\alpha}^{\dagger} C_{j\alpha}=1$
for all $j$), the supersymmetric exchange operator is defined as 
\beq
\hat{P}_{ij}^{(m|n)} \equiv
\sum_{\alpha,\beta=1}^{m+n} C_{i \alpha}^\dagger
C_{j \beta}^\dagger C_{i \beta}C_{j \alpha} \, . 
\label{b7b}
\eeq
This supersymmetric exchange operator
can equivalently be described as an operator on a spin space 
in the following way. Let us assume that each lattice site  
of a spin chain is occupied by either one of the $m$ number 
of `bosonic' spins or one of the $n$ number of `fermionic' spins. 
Hence, the total internal space associated with
such spin chain can be expressed as
\beq
\mbs{\Sigma}^{(m|n)} \equiv \underbrace{\mc{C}_{m+n} 
\otimes \mc{C}_{m+n} 
\otimes \cdots \otimes \mc{C}_{m+n}}_{N} \, . 
\label{b8}
\eeq
Using the notation of Ref.~\cite{BFGR09}, 
the orthonormal basis vectors of $\mbs{\Sigma}^{(m|n)}$  
may be denoted as $\ket{s_1,\cdots,s_N}$, where 
$s_i \equiv (s_i^1,s_i^2)$ is a vector with two components 
taking values within the range 
\begin{subequations}
\bea
&&s_i^1\equiv \pi(s_i) = \left \{  
\begin{array}{ll} 
0,  & \mbox{~for bosons, } \\
1, &   \mbox{~for fermions, } 
\end{array} 
\right.
\label{b9a}\\
&&s_i^2 \in \left \{ 
\hskip -.4 cm 
\begin{array}{ll} 
&\mbox{~ $\{-\frac{m-1}{2},-\frac{m-1}{2}+1,
\cdots,\frac{m-1}{2}\},~~\mr{if} ~\pi(s_i)=0$, } \\
&~\mbox{~$\{-\frac{n-1}{2},-\frac{n-1}{2}+1,
\cdots,\frac{n-1}{2}\},~~\mr{if} ~\pi(s_i)=1$. } 
\end{array} 
\right.
\label{b9b}
\eea
\label{b9}
\end{subequations}
Thus the component $s_i^1\equiv \pi(s_i)$ denotes the 
type of spin (bosonic or fermionic) and the component
$s_i^2$ denotes the numerical value of the spin.
A supersymmetric spin exchange 
operator $P_{ij}^{(m|n)}$ has been defined earlier 
on the space $\mbs{\Sigma}^{(m|n)}$ as~\cite{BB06,Ba99} 
\beq
P_{ij}^{(m|n)}\ket{s_1,\cdots,s_i,\cdots,s_j,\cdots,s_N}
=(-1)^{\alpha_{ij}(\mbf{s})}
\ket{s_1,\cdots,s_j,\cdots,s_i,\cdots,s_N},
\label{b10}
\eeq
where
$\alpha_{ij}(\mbf{s})
=\pi(s_i)\pi(s_j)+\left(\pi(s_i)+\pi(s_j)\right)\, 
h_{ij}(\mbf{s})$
and
$h_{ij}(\mbf{s})=\sum_{k=i+1}^{j-1}\pi(s_k)$ denotes the 
number of fermions in between the $i$-th and $j$-th spins.
From \Eq{b10} it follows that, 
the exchange of two bosonic (fermionic) spins produces
a phase factor of $1 (-1)$.
However, the exchange one bosonic spin with one fermionic
spin (or, vice versa) produces a phase factor of 
$(-1)^{h_{ij}(\mbf{s})}$. 
Using the commutation (anti-commutation) relations in \eq{b7a}, 
it can be shown that $\hat{P}_{ij}^{(m|n)}$
in \eq{b7b} is completely equivalent to  
$P_{ij}^{(m|n)}$ in \eq{b10}~\cite{BB06,Ba99}. 

A supersymmetric analogue of the spin reversal operator $P_i$ \eq{b4}
can also be defined on the space 
$\mbs{\Sigma}^{(m|n)}$~\cite{BFGR09}.
While acting on the basis vectors of $\mbs{\Sigma}^{(m|n)}$,
this supersymmetric analogue of spin reversal operator
(denoted by $P_i^{\ep\ep^{\prime}}$)
reverses the value of 
the $i$-th spin without affecting its type and multiplies 
the state by a sign factor. More precisely, the action of 
$P_i^{\ep\ep^{\prime}}$ is given by 
\beq
P_i^{\ep\ep^{\prime}}\ket{s_1,\cdots,s_i,\cdots,s_N}
=\rho(s_i)\ket{s_1,\cdots,s_i^{-},\cdots,s_N},
\label{b11}
\eeq
where $s_i^{-}=(s_i^1,-s_i^2)$, $\rho(s_i)= \ep ~(\ep^{\prime}) $
for $\pi(s_i)=0 ~(1)$, 
and $\ep,\ep^{\prime}=\pm 1$ are two independent signs.
With the help of \eq{b10} and \eq{b11}, one can easily 
check that $P_{ij}^{(m|n)}$ and $P_i^{\ep\ep^{\prime}}$  
yield a realization of the $BC_N$ type of Weyl algebra \eq{b1}.
Substitution of $P_{ij}^{(m|n)}$ and $P_i^{\ep\ep^{\prime}}$  
in \Eq{b1a} yields an exactly solvable 
Hamiltonian given by~\cite{BFGR09}
\beq
\mathcal{H}^{(m|n)}_{\ep \ep'}
=\sum_{i\neq j}\left[\frac{1-P_{ij}^{(m|n)}}{(\xi_i-\xi_j)^2} +
\frac{1-\widetilde{P}_{ij}^{(m|n)}}{(\xi_i+\xi_j)^2}\right]
+\beta\sum_{i=1}^{N}\frac{1-P_i^{\ep \ep'}}{\xi_i^2} \, , 
\label{b11a}
\eeq
where $\widetilde{P}_{ij}^{(m|n)}= P_i^{\ep \ep'}P_j^{\ep \ep'}
P_{ij}^{(m|n)}$. However, since $\mathcal{H}^{(m|n)}_{\ep \ep'}$
in the above equation does not reduce to  
${\mc{H}}^{(m,0)}$ in \eq{a2} 
for the special case $n=0$ (and for any possible choice of 
$\ep$ and $\ep^{\prime}$),
the former Hamiltonian can not be considered
as a supersymmetric extension of the later one. 

At present our aim is to construct SAPSRO 
which would satisfy the $BC_N$ type of Weyl algebra \eq{b1}. 
To this end, 
we denote the total internal space of the related spin system 
as $\mbs{\Sigma}^{(m_1,m_2|n_1,n_2)}$, 
where $m_1,~m_2,~n_1,~ n_2$ are some arbitrary non-negative integers
satisfying the relations $m_1+m_2=m$ and $n_1+n_2=n$. 
This $\mbs{\Sigma}^{(m_1,m_2|n_1,n_2)}$
can be expressed in a direct product form exactly like   
\eq{b8}, but each $s_i$ within the corresponding basis vectors  
now possess an extra quantum number
associated with the action of SAPSRO. 
More precisely, $\mbs{\Sigma}^{(m_1,m_2|n_1,n_2)}$ is spanned 
by orthonormal state vectors like  $\ket{s_1,\cdots,s_N}$, where 
$s_i \equiv (s_i^1,s_i^2,s_i^3)$ is a vector with three components 
taking values within the range 
\begin{subequations}
\bea
&&s_i^1\equiv \pi(s_i) = \left \{  
\begin{array}{ll} 
0,  & \mbox{~for bosons, } \\
1, &   \mbox{~for fermions, } 
\end{array} 
\right.
\label{b12a}\\
&&s_i^2\equiv f(s_i) = \left \{  
\begin{array}{ll} 
0,  & \mbox{~for positive parity under SAPSRO} \\
1, &   \mbox{~for negative parity under SAPSRO, } 
\end{array} 
\right.
\label{b12b}\\
&&s_i^3 \in \left \{ \hskip -.47 cm  
\begin{array}{llll} 
&\mbox{~ $\{1,2, \cdots,m_1 \},
~\mr{if}~\pi(s_i)=0~ \mr{and} ~f(s_i)=0$, } \\
&~\mbox{~$\{1, 2, \cdots, m_2 \},
~\mr{if}~\pi(s_i)=0 ~ \mr{and} ~ f(s_i)=1 $, } \\
&~\mbox{~$\{1, 2, \cdots, n_1 \},
~\mr{if}~\pi(s_i)=1~ \mr{and} ~f(s_i)=0 $, } \\
&~\mbox{~$\{1, 2, \cdots, n_2\},
~\mr{if}~\pi(s_i)=1~ \mr{and} ~f(s_i)=1 $. } 
\end{array} 
\right.
\label{b12c}
\eea
\label{b12}
\end{subequations}
Indeed, we define the action of SAPSRO 
(denoted by $P_i^{(m_1,m_2|n_1,n_2)}$) on these state vectors as 
\beq
P_i^{(m_1,m_2|n_1,n_2)}\ket{s_1,\cdots,s_i,\cdots, s_N}
=(-1)^{f(s_i)}\ket{s_1,\cdots,s_i,\cdots, s_N},
\label{b13}
\eeq
which shows that $s_i^2\equiv f(s_i)$ 
is determined through the parity of the spin 
$s_i$ under the action of SAPSRO. As before, 
the action of supersymmetric spin exchange operator 
$P_{ij}^{(m|n)}$ on the space $\mbs{\Sigma}^{(m_1,m_2|n_1,n_2)}$
is given by an equation of the form \eq{b10}, where   
the phase factor $\alpha_{ij}(\mbf{s})$
depends on the first components of the spins   
like $s_k^1\equiv \pi(s_k)$. 
Using Eqs.~\eq{b10} and \eq{b13}, we find that 
$P_{ij}^{(m|n)}$ and $P_i^{(m_1,m_2|n_1,n_2)}$  
yield a realization of the $BC_N$ type of Weyl algebra \eq{b1}.
Substituting these operators   
in the general form of Hamiltonian \eq{b1a}, we obtain 
the Hamiltonian for a large 
class of $BC_N$ type of PF spin chains as 
\beq
\mathcal{H}^{(m_1,m_2|n_1,n_2)}
=\sum_{i\neq j}\left[\frac{1- P_{ij}^{(m|n)}}{(\xi_i-\xi_j)^2} +
\frac{1- \widetilde{{P}}_{ij}^{(m_1,m_2|n_1,n_2)}}{(\xi_i+\xi_j)^2}\right]
+\beta\sum_{i=1}^{N}\frac{1-P_i^{(m_1,m_2|n_1,n_2)} }{\xi_i^2} \, ,  
\label{b14}
\eeq
where $\widetilde{P}_{ij}^{(m_1,m_2|n_1,n_2)} \equiv 
 P_i^{(m_1,m_2|n_1,n_2)}P_j^{(m_1,m_2|n_1,n_2)} P_{ij}^{(m|n)}$.
 
It is worth noting that the Hamiltonian \eq{b14}
can reproduce all of the previously studied $BC_N$ type of 
PF spin chains at certain limits. For example, 
in the presence of only bosonic or fermionic spins, i.e., when 
  either $n_1=n_2=0$ or $m_1=m_2=0$, 
$\mathcal{H}^{(m_1,m_2|n_1,n_2)}$ 
reduces to the non-supersymmetric PF spin chain 
associated with PSRO~\cite{BBB14}. 
Next, let us assume that the discrete parameters 
$m_1,~m_2,~n_1,~ n_2$ in the Hamiltonian \eq{b14} 
satisfy the relations 
\beq
m_1= \frac{1}{2}\left( m+ \ep \, \tilde{m} \right),~ 
m_2= \frac{1}{2}\left( m - \ep \, \tilde{m} \right),~
n_1= \frac{1}{2}\left( n+ \ep' \, \tilde{n} \right), ~
n_2= \frac{1}{2}\left( n - \ep' \, \tilde{n} \right),
\label{b15}
\eeq
where $\ep,\ep^{\prime}=\pm 1$, $\tilde{m} \equiv m~ \mr{mod}~ 2$ and
$\tilde{n} \equiv n~ \mr{mod}~ 2$. One can easily check that, for these 
particular values of the discrete parameters, the trace 
of $P_i^{(m_1,m_2|n_1,n_2)}$ in \eq{b13} would 
coincide with that of $P_i^{\ep\ep^{\prime}}$
in \eq{b11}. 
Furthermore, it would be possible to construct 
an unitary transformation which maps $P_i^{(m_1,m_2|n_1,n_2)}$
to  $P_i^{\ep\ep^{\prime}}$ and keeps $P_{ij}^{(m|n)}$ invariant. 
Consequently, for the special case  
given in \eq{b15}, 
$\mathcal{H}^{(m_1,m_2|n_1,n_2)}$ in \eq{b14}
becomes equivalent to the exactly solvable Hamiltonian 
$\mathcal{H}^{(m|n)}_{\ep \ep'}$ in \eq{b11a}.

Except for the two particular cases
which are discussed above,
the Hamiltonian in \eq{b14} represents 
novel class of $BC_N$ type of PF spin chains
 associated with SAPSRO. 
For example, if we choose the discrete parameters 
as $m_1=m, ~m_2=0,~n_1=n, ~n_2=0$, 
then Eqs.~\eq{b12c} and \eq{b13} imply that $P_i^{(m,0|n,0)}=\one $ and   
$\widetilde{P}_{ij}^{(m,0|n,0)} = P_{ij}^{(m|n)}$.
Consequently, for this particular case,  
$\mathcal{H}^{(m_1,m_2|n_1,n_2)}$ in  
\eq{b14} yields a 
supersymmetric spin chain of the form 
\beq
\mathcal{H}^{(m,0|n,0)}
=\sum_{ i \neq j }
 \frac{y_i+y_j}{(y_i-y_j)^2} \left( 1- P_{ij}^{(m|n)} \right) \, , 
 \label{b16}
\eeq
which has not been studied previously in the literature. 
It is interesting to observe that, for the special case $n=0$,
the above Hamiltonian 
reduces to ${\mc{H}}^{(m,0)}$ in \eq{a2} with $\ep=1$.
On the other hand, by putting $n=0$ after interchanging 
$m$ and $n$ in \eq{b16}, one easily gets ${\mc{H}}^{(m,0)}$ with $\ep=-1$.
Therefore, the Hamiltonian $\mathcal{H}^{(m,0|n,0)}$
in \eq{b16}  can be considered
as a supersymmetric extension of ${\mc{H}}^{(m,0)}$ in \eq{a2}. 

We would like to make a comment at this point. 
The integrability of the Hamiltonian 
$\mathcal{H}^{(m_1,m_2|n_1,n_2)}$ in  
\eq{b14} can  be established  by using a procedure 
similar to that of Ref.~\cite{YT96}
in the non-supersymmetric case.
However, there exists an important difference between the 
symmetry algebra of spin chains
associated with the $BC_N$ root system
and that of spin chains associated with the $A_{N-1}$ root system.
As is well known, the Hamiltonian
\eq{a1} of the $A_{N-1}$ type of PF spin chain exhibit 
global \su{m} symmetry
along with more general $Y(gl(m))$ 
Yangian quantum group symmetry \cite{Hi95npb}. 
Moreover, the supersymmetric extension of this 
$A_{N-1}$ type of PF spin
exhibit global \su{m|n} supersymmetry as well as  
$Y(gl(m|n))$ super Yangian symmetry \cite{HB00}. On the other
hand, PF spin chains 
associated with the $BC_N$ root system do not, in general, 
exhibit global \su{m} symmetry or \su{m|n} supersymmetry.
For example, the presently considered Hamiltonian 
$\mathcal{H}^{(m_1,m_2|n_1,n_2)}$ in  
\eq{b14}, which depends on operators like $P_{ij}^{(m|n)}$
and $P_i^{(m_1,m_2|n_1,n_2)}$, 
does not commute with all generators of the    
\su{m|n} super Lie algebra for arbitrary values of 
the discrete parameters $m_1,~m_2,~n_1$ and $n_2$. This happens 
because, while $P_{ij}^{(m|n)}$ 
commutes with all generators of the \su{m|n} super
Lie algebra, $P_i^{(m_1,m_2|n_1,n_2)}$ defined in \eq{b13} does not  
commute with those generators  
for arbitrary values of the discrete parameters. However, 
we have already mentioned that in the particular case given by 
$m_1=m, ~m_2=0,~n_1=n, ~n_2=0$, 
$P_i^{(m_1,m_2|n_1,n_2)}$ reduces to the trivial 
identity operator. Consequently, the corresponding
Hamiltonian $\mathcal{H}^{(m,0|n,0)}$
in \eq{b16} commutes with all generators of the \su{m|n} super
Lie algebra.


\noi \section{Spectra and partition functions of 
$BC_N$ type models with SAPSRO }
\renewcommand{\theequation}{3.{\arabic{equation}}}
\setcounter{equation}{0}
\medskip

In the following, our aim is to compute 
 the partition functions of the $BC_N$ type of PF spin 
chains \eq{b14} for all possible choice of the corresponding discrete
parameters.
To this end, we shall consider   
a class of $BC_N$ type of spin Calogero models with SAPSRO 
and, by using the freezing trick, 
show that the strong coupling limit 
of such spin Calogero models leads to  
the Hamiltonian $\mathcal{H}^{(m_1,m_2|n_1,n_2)}$ in \eq{b14}. 
Next, we shall find out 
the exact spectra for the above mentioned  
$BC_N$ type of spin Calogero models with SAPSRO 
and also compute the corresponding partition functions in the 
strong coupling limit. 
Finally, by `modding out' the contribution of the 
coordinate degrees of freedom from the 
above mentioned partition functions, 
we shall obtain an exact expression
for the partition functions of the $BC_N$ type of PF spin 
chains \eq{b14}. 

By using SAPSRO in \eq{b13}, 
let us define 
the Hamiltonian for a class of $BC_N$ type of 
spin Calogero models as
\bea
H^{(m_1,m_2|n_1,n_2)}=-\sum_{i=1}^{N}\frac{\d^2}{\d x_i^2} 
+\frac{a^2}{4} r^2
+a\sum_{i\neq j} \left[\frac{a-P_{ij}^{(m|n)}}
{(x_{ij}^-)^2} +
\frac{a-\widetilde{P}_{ij}^{(m_1,m_2|n_1,n_2)}}{(x_{ij}^+)^2}\right] \nn \\
+\beta a\sum_{i=1}^{N}\frac{\beta a -P_i^{(m_1,m_2|n_1,n_2)}}{x_i^2}\, ,
\label{c1}
\eea
where $a> \frac{1}{2}, ~\beta>0$ are real coupling constants and  
the notations $x_{ij}^- \equiv x_i-x_j$, $x_{ij}^+ \equiv x_i+x_j$, 
$r^2\equiv \sum_{i=1}^N x_i^2$ are used. It should be noted that 
this Hamiltonian 
contains both coordinate and spin degrees of freedom.
Similar to the case of $BC_N$ type of 
spin Calogero models considered earlier~\cite{YT96,BFGR08,BFGR09,BBB14}, 
the potentials of $H^{(m_1,m_2|n_1,n_2)}$ in 
\eq{c1} become singular in the limits $x_i \pm x_j \to 0$ and 
$x_i \to 0$. Therefore, the configuration space of this Hamiltonian 
can be taken as one of the maximal open subsets of $\mbb{R}^N$ on which 
linear functionals $x_i\pm x_j$ and $x_i$ have constant signs.  
Let us choose this configuration space as the principal 
Weyl chamber of the $BC_N$ root system given by
\beq
 C = \{ \mbf{x} \equiv (x_1,x_2,\cdots ,x_N) 
 :~ 0<x_1<x_2<\ldots<x_N \} \, .
\label{c2}
\eeq 
Next, we express $H^{(m_1,m_2|n_1,n_2)}$ \eq{c1} 
in powers of the coupling constant $a$
as 
\beq
H^{(m_1,m_2|n_1,n_2)} = 
-\sum_{i=1}^N \frac{\partial^2}{ \partial x_i^2 }
+ a^2 \, U(\mbf{x}) + O(a) \, ,
\label{c3}
\eeq 
with 
\beq
U(\mbf{x})=
\sum_{i\neq j}\left[\frac{1}{(x_{ij}^-)^2}+\frac{1}{(x_{ij}^+)^2}\right]
+\beta^2\sum_{i=1}^{N}\frac{1}{x_i^2}+\frac{r^2}{4}.
\label{c4}
\eeq
Since the $a^2$ order term in \eq{c3} dominates 
in the strong coupling limit $a\to \infty $,
the particles of $H^{(m_1,m_2|n_1,n_2)}$ concentrate at the coordinates 
$\xi_i$ of the minimum $\mbs{\xi}$ of the potential $U(\mbf{x})$ in $C$.
As a result, 
the coordinate and spin degrees of freedom 
of these particles 
decouple from each other and the Hamiltonian
$H^{(m_1,m_2|n_1,n_2)}$ in \eq{c1} can be written
in $a\to \infty$ limit as 
\beq
H^{(m_1,m_2|n_1,n_2)}
\approx
H_{sc}+a\, \mathfrak{H}^{(m_1,m_2|n_1,n_2)}|_{\mbf{x} \to \mbs{\xi}}
\, ,
\label{c5}
\eeq
where $H_{sc}$ is the scalar (spinless) Calogero model 
of $BC_N$ type given by 
\beq
H_{sc}=-\sum_{i=1}^N \frac{\partial^2}{\partial x_i^2}
+\frac{a^2}{4}\,r^2
+a(a-1)\sum_{i\neq 
j}\bigg[
\frac{1}{(x_{ij}^-)^2}+\frac{1}{(x_{ij}^+)^2}\bigg]
+\sum_{i=1}^N\frac{a\beta (a \beta -1)}{x_i^2}\,,
\label{c6}
\eeq 
and 
\beq
\mathfrak{H}^{(m_1,m_2|n_1,n_2)}
=\sum_{i\neq j}\left[\frac{1-P_{ij}^{(m|n)}}{(x_i-x_j)^2} +
\frac{1-\widetilde{P}_{ij}^{(m_1,m_2|n_1,n_2)}}{(x_i+x_j)^2}\right]
+\beta\sum_{i=1}^{N}\frac{1-P_i^{(m_1,m_2|n_1,n_2)}}{x_i^2} \, .
\label{c7}
\eeq
The uniqueness of the unique minimum $\mbs{\xi}$ 
of the potential $U$ 
\eq{c4} within the configuration space $C$ \eq{c2}
has been established in Ref.~\cite{CS02}
by expressing this potential
in terms of the logarithm of the ground state wave function  
of the scalar Calogero model \eq{c6}.   
The ground state wave function  
of this scalar Calogero model, 
with ground state energy  
\beq
\label{E0}
E_0=Na\Big(\beta a +a(N-1)+\frac{1}{2} \Big)\, , 
\eeq
is given by   
\beq
\mu({\bf x})=e^{-\frac{a}{4}r^2}~\prod_i|x_i|^{\beta a}
~\prod_{i<j}|x_i^2-x_j^2|^a \, . 
\label{c9}
\eeq
Using the fact that the sites $\xi_i$ coincide with the 
coordinates of the (unique) critical point of 
$\log \mu({\bf x})$ in $C$, 
one obtains a set of relations 
among these sites as 
\cite{CS02,BFGR08}
\beq
\sum_{\stackrel{j=1}{(j\neq i)}}^N \, 
\frac{2y_i}{y_i-y_j} \, = \, y_i-\be \, ,
\label{c10}
\eeq
where $\xi_i=\sqrt{2y_i}$ and $y_i$'s  
denote the zeros of the generalized Laguerre polynomial 
$L_N^{\beta -1}$. Consequently, the operator 
$\mathfrak{H}^{(m_1,m_2|n_1,n_2)}|_{\mbf{x} \to \mbs{\xi}}$ in 
\eq{c5} coincides with the 
Hamiltonian $\mathcal{H}^{(m_1,m_2|n_1,n_2)}$ \eq{b14}
of PF spin chains with SAPSRO.
Furthermore, due to \Eq{c5}, 
eigenvalues of ${H}^{(m_1,m_2|n_1,n_2)}$ are approximately given by 
\beq
E_{ij}^{(m_1,m_2|n_1,n_2)} 
\simeq E_i^{sc} + a \, \mc{E}_j^{(m_1,m_2|n_1,n_2)} \, ,
\label{c12}
\eeq
where $E_i^{sc}$ and $\mc{E}_j^{(m_1,m_2|n_1,n_2)}$ 
are two arbitrary eigenvalues
of $H_{sc}$ and $\mathcal{H}^{(m_1,m_2|n_1,n_2)}$ respectively. 
With the help of \Eq{c12}, we obtain an exact formula 
for the partition function
$\mc{Z}_N^{(m_1,m_2|n_1,n_2)}(T)$ of the spin chain \eq{b14} 
at a given temperature $T$ as 
\beq
\mc{Z}_N^{(m_1,m_2|n_1,n_2)}(T)=\lim_{a \rightarrow \infty}
 \frac{Z^{(m_1,m_2|n_1,n_2)}_N(aT)}{Z_N(aT)} \, ,
\label{c13}
\eeq
where $Z_N^{(m_1,m_2|n_1,n_2)}(T)$ represents
the partition function 
of the $BC_N$ type of spin Calogero Hamiltonian (\ref{c1}) and
$Z_N(T)$ represents that of the scalar model (\ref{c6}).

An exact expression for the partition function of the  
scalar model (\ref{c6}) has been obtained 
earlier as~\cite{BFGR08} 
\beq
Z_N(aT)=
\frac{q^{\frac{{E}_0}{a}}}{\prod\limits_{j=1}^N (1-q^{2j})} \, ,
\label{c22}
\eeq 
where $q=e^{-1/(k_BT)}$.
Therefore, for the purpose of  
evaluating the partition function 
$\mc{Z}_N^{(m_1,m_2|n_1,n_2)}(T)$ of the spin chain \eq{b14}
by using \Eq{c13}, it is required to 
compute the spectrum and partition function of spin Calogero 
Hamiltonian $H^{(m_1,m_2|n_1,n_2)}$ in \eq{c1}.
To this end, we start with the  
$BC_N$ type of auxiliary operator
given by~\cite{BFGR08}
\beq
\mbb{H}=-\sum_{i=1}^N \frac{\partial^2}{ \partial x_i^2 }
+a \, \sum_{i\neq j}  \left[
 \frac{a-K_{ij}}{(x_{ij}^-)^2}+ \frac{a-\widetilde{K}_{ij}}{(x_{ij}^+)^2}\right]
+\beta a \, \sum_{i=1}^N \frac{\beta a-K_i}{x_i^2}+\frac{a^2}4\,r^2\,,
\label{auxi}
\eeq
where   
$K_{ij}$ and $K_i$ are coordinate permutation and sign reversing
operators, defined by
\bse
\bea
&& (K_{ij}f)(x_1,\dots,x_i,\dots,x_j,\dots,x_N)=f(x_1,\dots,x_j,\dots,x_i,
\dots,x_N)\,,  \label{action1}\\
&& (K_i f)(x_1,\dots,x_i,\dots,x_N)=f(x_1,\dots,-x_i,\dots,x_N)\,, 
\label{action2} 
\eea 
\label{action}
\ese
and $\widetilde{K}_{ij}=K_iK_jK_{ij}$. 
As shown in the latter reference, 
the auxiliary operator \eq{auxi}
can be written as 
\beq
\mathbb{H}= \mu(\mbf{x}) \left[ -\sum_i  \big(J_i\big)^2+a \sum_i x_i 
\frac{\partial}{\partial x_i}
+ E_0 \right] \mu^{-1}(\mbf{x})\, ,  
\label{c15}
\eeq
where $J_i$'s are $BC_N$ type of Dunkl operators 
given by
\bea
J_i=\frac{\partial}{\partial x_i} +a \sum_{j\neq i}
\left[ \frac{1}{x_{ij}^-}
(1-K_{ij})\,
+\frac{1}{x_{ij}^+}(1-\widetilde{K}_{ij})\right] 
+\beta a\, \frac{1}{x_i}(1-K_i)\, ,
\label{c14}
\eea
with $i \in \{ 1,2, \dots,N \}$. 
Let us now consider a Hilbert space spanned by a set of
 basis vectors like 
\beq
\phi_{\bf{r}}({\bf x})=\mu({\bf x})\prod_ix_i^{r_i} ,
\label{c16}
\eeq
with $r_i$'s being arbitrary non-negative integers,  
and (partially) order these basis 
vectors according to their total degree
$|\mbf{r}| \equiv r_1+r_2+\cdots + r_N $. Since
the Dunkl operators~\eq{c14} clearly map any monomial 
$\prod_i x_i^{r_i}$ into a
polynomial of total degree $r_1+ r_2+\cdots+r_N-1$,
it follows from \Eq{c15}  that  
$\mathbb{H}$  acts as an upper triangular matrix in the
aforementioned non-orthonormal basis:  
\beq
\mathbb{H} \phi_{\bf r}({\bf x})=E_{\mbf{r}} \phi_{\bf r}({\bf x})+
\sum_{\vert { \bf r'} \vert < \vert {\bf r} \vert }c_{{\bf r'} {\bf r} }\,
\phi_{\bf r'} ({\bf x})\,,
\label{c17}
\eeq
where
\begin{equation}\label{Ep}
E_{\bf r}=a \vert  {\bf r}  \vert+E_0 \,  ,
\end{equation}
and the coefficients $c_{{\bf r' } {\bf r} }$ are
some real constants.
Hence the spectrum of $\mbb{H}$ is given by
the diagonal entries of this upper triangular matrix, i.e.,   
$E_{\bf r}$'s in \Eq{Ep},   
where $r_i$'s can be taken as arbitrary non-negative integers.

In the following, we shall compute 
the spectrum of the spin Calogero 
Hamiltonian $H^{(m_1,m_2|n_1,n_2)}$ 
from that of $\mbb{H}$ 
by taking advantage of the fact that these two operators are 
related through formal substitutions like 
\beq
H^{(m_1,m_2|n_1,n_2)}= \mathbb{H}|_{K_{ij} \rightarrow
P_{ij},\, K_i \rightarrow P_i^{(m_1,m_2|n_1,n_2)}} \,  .
\label{c18} 
\eeq
Due to the impenetrable nature of the singularities of the
spin Calogero Hamiltonian $H^{(m_1,m_2|n_1,n_2)}$, 
its Hilbert space can be taken as the space 
$L^2(C)\otimes \mbs{\Sigma}^{(m_1,m_2|n_1,n_2)}$
of wave functions square integrable on the set $C$ 
in \Eq{c2}. However,
any point in $\mbb{R}^N$ not lying within the 
singular subset $x_i\pm x_j=0$, $x_i=0$, $1\le i<j\le N$, 
can be mapped in a unique way to a point in $C$ 
by an element of the $BC_N$ Weyl group~\cite{Hu90}. 
Using this fact, 
it can be shown that 
$L^2(C)\otimes \mbs{\Sigma}^{(m_1,m_2|n_1,n_2)}$
is isomorphic to the Hilbert space $\mbb{V}$ defined as  
\beq
\mbb{V} \equiv \Lambda^{(m_1,m_2|n_1,n_2)}(L^2(\mbb{R}^N)\otimes
 \mbs{\Sigma}^{(m_1,m_2|n_1,n_2)} ) \, ,   
\label{vsp}
\eeq 
with 
$\Lambda^{(m_1,m_2|n_1,n_2)}$ being a projector which 
satisfies the relations 
\begin{subequations}
\bea
 & \Pi_{ij}^{(m|n)} \, \Lambda^{(m_1,m_2|n_1,n_2)} =
\Lambda^{(m_1,m_2|n_1,n_2)} \, \Pi_{ij}^{(m|n)}
= \, \Lambda^{(m_1,m_2|n_1,n_2)}  
\label{c23}, \\
& \Pi_i^{(m_1,m_2|n_1,n_2)}\, \Lambda^{(m_1,m_2|n_1,n_2)}
=\Lambda^{(m_1,m_2|n_1,n_2)} \, \Pi_i^{(m_1,m_2|n_1,n_2)}
=\Lambda^{(m_1,m_2|n_1,n_2)}, 
\label{c24}
\eea
\label{c25}
\end{subequations}
where $\Pi_{ij}^{(m|n)} \equiv K_{ij}P_{ij}^{(m|n)}$ and 
$ \Pi_i^{(m_1,m_2|n_1,n_2)} \equiv  K_iP_i^{(m_1,m_2|n_1,n_2)}$.
Following the usual procedure of constructing projectors
associated with the $BC_N$ type of 
Weyl algebra~\cite{FGGRZ03,CC04},
we obtain an expression for $\Lambda^{(m_1,m_2|n_1,n_2)}$ 
satisfying \eq{c25} as 
\beq
\Lambda^{(m_1,m_2|n_1,n_2)} =
\frac{1}{2^N \cdot N!}\left\{\prod_{j=1}^N
\Big(1+   \Pi_j^{(m_1,m_2|n_1,n_2)}  \Big)
\right\} \sum_{l=1}^{N!}\, \mc{P}_l \, , 
\label{pro22}
\eeq
where $\mc{P}_l$ denotes the realization of an element 
of the permutation group (for $N$ number of particles)
through the operators $\Pi_{ij}^{(m|n)}$. 
For example, in the simplest $N=2$ case, 
\Eq{pro22} yields 
\[ \Lambda^{(m_1,m_2|n_1,n_2)} =\frac{1}{8}
\Big(1+ \Pi_1^{(m_1,m_2|n_1,n_2)} \Big)
\Big(1+\Pi_2^{(m_1,m_2|n_1,n_2)}\Big)
(1+\Pi_{12}^{(m|n)}). \]
It may be noted that 
 $\Lambda^{(m_1,m_2|n_1,n_2)}$  in \eq{pro22} 
 commutes with the auxiliary operator in \eq{auxi}:
\beq
 \left[\Lambda^{(m_1,m_2|n_1,n_2)}, \mbb{H} \right ] = 0 \, .
\label{comm}
\eeq
Since $H^{(m_1,m_2|n_1,n_2)}$
is equivalent to its natural extension to the space $\mbb{V}$ \eq{vsp},
with a slight abuse of notation we also denote 
the latter operator as $H^{(m_1,m_2|n_1,n_2)}$. 
Thus, by using the relations \eq{c25}, we
can transform Eq.~\eq{c18}  
into an operator relation given by  
\begin{equation}\label{HHp}
H^{(m_1,m_2|n_1,n_2)} \Lambda^{(m_1,m_2|n_1,n_2)}=
\mbb{H} \Lambda^{(m_1,m_2|n_1,n_2)} \, .  
\end{equation}

We shall now explain how the operator relation \eq{HHp} 
plays an important role in finding the 
spectrum of $H^{(m_1,m_2|n_1,n_2)}$ from that of $\mbb{H}$.
To this end, it may be noted that the Hilbert space $\mbb{V}$
in \eq{vsp} is the closure of the linear subspace spanned by the 
wave functions of the form 
\beq
\psi_{\bf{r}}^{\bf{s}} 
\equiv 
\psi_{ r_1 ,\ldots, r_i, \ldots, r_j, \ldots , r_N}
^{s_1, \ldots, s_i, \ldots, s_j, \ldots, s_N}
=\Lambda^{(m_1,m_2|n_1,n_2)}
\left (\phi_{\bf{r}} ({\bf x})|{\bf s}\rangle \right) \, ,
\label{c26}
\eeq
where $\phi_{\bf{r}}$ is given in \eq{c16} and 
$\ket{\mbf{s}} \equiv  \ket{s_1,\cdots,s_N}$
is an arbitrary basis element of 
the spin space  $\mbs{\Sigma}^{(m_1,m_2|n_1,n_2)}$.
However, $\psi_{\bf{r}}^{\bf{s}}$'s 
defined in \Eq{c26} do not form a set of 
linearly independent state vectors.
Indeed, by using \eq{c23},  \eq{action1} and  
an equation of the form \eq{b10} for the basis elements of 
$\mbs{\Sigma}^{(m_1,m_2|n_1,n_2)}$, we find that 
$\psi_{\bf{r}}^{\bf{s}}$'s 
satisfy the condition  
\beq
\psi_{ r_1 ,\ldots, r_i, \ldots, r_j, \ldots , r_N}
^{s_1, \ldots, s_i, \ldots, s_j, \ldots, s_N}
=(-1)^{\alpha_{ij}(\mbf{s})} \, 
\psi_{r_1 , \ldots, r_j, \ldots, r_i, \ldots, r_N}^
{s_1, \ldots, s_j, \ldots, s_i, \ldots , s_N} \, . 
\label{c27}
\eeq
Moreover, by using \eq{c24}, 
\eq{action2} and \eq{b13}, we obtain 
\beq
\psi_{r_1, \ldots, r_N}^{s_1, \ldots, s_N}=(-1)^{r_i+f(s_i)}~\psi_{r_1,
\ldots, r_N}^{s_1, \ldots, s_N} \, .
\label{c28}
\eeq
Due to Eqs.~\eq{c27} and \eq{c28} it follows that, 
$\psi_{\bf{r}}^{\bf{s}}$'s 
defined through \Eq{c26} would be nontrivial and 
linearly independent if 
the following three conditions are imposed on the corresponding  
$r_i$'s and $s_i$'s.

1) An ordered form of $\mbf{r}$, 
which separately arranges its even and odd components 
into two non-increasing sequences, i.e.,   
\bea
{\mbf r}~\equiv ~(\mbf{r_e}, \mbf {r_o})&=&(\overbrace{2l_1,
\ldots, 2l_1}^{k_1}, \,  \ldots, \,  \overbrace{2l_s, \ldots,
2l_s}^{k_s}, \nn \\ &~& \overbrace{2p_1+1, \ldots, 2p_1+1}^{g_1}, \, 
\ldots, \, \overbrace{2p_t+1,
\ldots, 2p_t+1}^{g_t}) \, ,
\label{c29}
\eea
where  $0\leq s,\, t \leq N$, 
$l_1>l_2>\ldots>l_s\geqslant0$ and $p_1>p_2>\ldots>p_t\geqslant0$, 
is chosen as the lower index of $\psi_{\bf{r}}^{\bf{s}}$.
It may be noted that, 
 any given $\mbf{r}$ can be brought 
 in the ordered form \eq{c29}
through an appropriate permutation of its 
components. Therefore, as a consequence of \Eq{c27},
we can choose the ordered form \eq{c29} 
in the lower index of independent state vectors.
\vskip 2mm
2) 
Using \Eq{c28}, we find that 
the second component of $s_i$ corresponding to each $r_i$ 
is given by   
\beq
s_i^2\equiv f(s_i) =\left \{
\begin{array}{ll} 0 , & \mbox{for} ~r_i\in \mbf {r_e} \,
, \\
1 \, , &
\mbox{for}~ r_i\in \mbf {r_o} \, . 
\end{array} 
\right. 
\label{c30}
\eeq
\vskip 2mm
3) Let us consider the special case where $r_i=r_j$ for $i<j$. Then, 
due to the condition 2), the second components of the 
corresponding spins $s_i$ and $s_j$ must have the same value.
In this special case, we can further     
use Eq.~\eq{c27} along with the definition of  
$\alpha_{ij}(\mbf{s})$ which appears just after Eq.~\eq{b10},
and arrange the first components of 
 $s_i$ and $s_j$ 
(and also their third components in some cases) 
 associated with independent state vectors 
such that \newline 

\hspace{2cm} i)~ $\pi(s_i)\leqslant \pi(s_j)$  , \\
\vspace{.05cm}
\hspace{2.5cm} ii)~ $s_i^3 \geqslant s_j^3 +\pi(s_j)$ ,
if $\pi(s_i)=\pi(s_j)$.

\vskip 2mm
All linearly independent $\psi_{\bf{r}}^{\bf{s}}$'s \eq{c26}, 
satisfying the above
mentioned three conditions, may now be taken as a set of 
(non-orthonormal) basis
vectors for the Hilbert space $\mbb{V}$
in \eq{vsp}. Let us define a partial ordering among
these basis vectors as:
 $\psi_{\bf{r}}^{\bf{s}}>\psi_{\bf{r}'}^{ \bf{s}'} \, $, 
 if~$|\bf{r}|>|\bf{r}'|$. 
Applying the key relation \eq{HHp}
along with \eq{c26}, we obtain 
\[
 H^{(m_1,m_2|n_1,n_2)}\psi_{\bf{r}}^{\bf{s}}= 
 \Lambda^{(m_1,m_2|n_1,n_2)}\left(\left(
\mbb{H}  \phi_{\bf{r}} ({\bf x})\right) |{\bf s}\rangle 
\right) .
 \]
Using this equation as well as \eq{comm} and \eq{c17}, 
we find that $H^{(m_1,m_2|n_1,n_2)}$ in \eq{c1} acts  
on the above mentioned partially ordered basis vectors 
of $\mbb{V}$ as
\beq
H^{(m_1,m_2|n_1,n_2)}\, \psi_{\bf{r}}^{\bf{s}}=E_{\mbf{r}}^{\mbf{s}} \,
 \psi_{\bf{r}}^{\bf{s}} +\sum_{|\bf{{r}'}|<|\bf{r}|}~C_{\mbf{r'r}} \, 
\psi_{\bf{{r}'}}^{\bf{s'}} \, ,
\label{c31}
\eeq
where $C_{\mbf{r'r}}$'s are real constants,  
$\mbf{s}'$ is a suitable permutation of $\mbf{s}$ and 
\beq
E_{\mbf{r}}^{\mbf{s}} =a|{\bf{r}}|+E_0 \, . 
\label{c32}
\eeq  
Due to such upper  
triangular matrix form of $H^{(m_1,m_2|n_1,n_2)}$,
all eigenvalues of this   
Hamiltonian are given by \Eq{c32}, where the quantum number $\mbf{r}$ 
satisfies the condition 1) 
and the quantum number $\mbf{s}$ satisfies the conditions 2) and 3). 
Since the RHS of \Eq{c32} does not depend on the spin 
quantum number $\mbf{s}$, 
 the eigenvalue associated with the quantum number
$\mbf{r}$ in \Eq{c29}
has an {\em intrinsic degeneracy} 
$d^{\,(m_1,m_2|n_1,n_2)}_{\, \bf k, \bf g}$
which counts the number of all possible   
choice of corresponding
spin degrees of freedom. 
Using the conditions 2) and 3),
we compute this intrinsic spin degeneracy 
associated with the quantum number $\mbf{r}$ as 
\beq
d^{\,(m_1,m_2|n_1,n_2)}_{\, \bf k, \bf g}
=\prod_{i=1}^s d_{\, m_1,n_1}(k_i)\, \prod_{j=1}^t d_{m_2,n_2}(g_j),
\label{c33}
\eeq
where the function $d_{\, x,y}(\nu)$ is given by 
\beq
d_{\, x,y}(\nu)= \sum_{i=0}^{\min(\nu,y)}
\binom{y}{i}\binom{x+\nu-i-1}{\nu-i}\,. 
\label{c33a}
\eeq
Due to \Eq{c32}, the actual degeneracy of an energy $a E_1 + E_0$ is 
evidently obtained by summing over the intrinsic degeneracy \eq{c33}
for all multi-indices $\mbf{r}$ in \eq{c29} with
fixed order $E_1$. Consequently, the actual degeneracy factors  
for the energy levels of spin Calogero Hamiltonian
$H^{(m_1,m_2|n_1,n_2)}$
in \eq{c1} would depend on the discrete 
parameters $m_1$, $m_2$, $n_1$ and $n_2$.

Let us now calculate the partition function for the
Hamiltonian $H^{(m_1,m_2|n_1,n_2)}$. 
Since  $|\mbf{r}|$ corresponding to the 
multi-index $\mbf{r}$ in \eq{c29} is given by 
$2\sum_{i=1}^s l_i  k_i+ 2\sum_{j=1}^t p_j g_j + \sum_{j=1}^t g_j  $,
we can express the energy eigenvalues \eq{c32} 
of $H^{(m_1,m_2|n_1,n_2)}$ as  
\beq
E_{\mbf{r}}^{\mbf{s}} 
=2a\sum_{i=1}^s l_i  k_i+ 2a\sum_{j=1}^t p_j g_j +  a\sum_{j=1}^t g_j
+E_0 \, .
\label{c34}
\eeq
By using \Eq{c29}, we obtain 
the numbers of the even and the odd components 
of $\mbf{r}$ (denoted by $N_1$ and $N_2$ respectively) as
\beq
N_1= \sum_{i=1}^s  k_i, ~~~~ N_2 = \sum_{j=1}^t g_j\, ,  \nn
\eeq
which satisfy the condition $N_1 +N_2=N$. 
Hence, we can write  
${\bf k} \equiv\{k_1, k_2, \ldots, k_s\} \in \mc{P}_{N_1} $ 
and ${\bf g} \equiv \{g_1, g_2, \ldots, g_t\} \in \mc{P}_{N_2} $, 
where $\mc{P}_{N_1}$ and  $\mc{P}_{N_2}$
denote the sets of all ordered partitions 
of $N_1$ and $N_2$ respectively.
Next, we compute the sum over the Boltzmann weights 
corresponding to all $\mbf{r}$'s of the form \eq{c29} 
with energy eigenvalues  \eq{c34} and 
intrinsic degeneracy factors  \eq{c33}.
Thus, we obtain the canonical partition function for the
$BC_N$ type of spin Calogero model  
\eq{c1} with SAPSRO as
\beq
Z_N^{(m_1,m_2|n_1,n_2)}(aT) =q^{\frac{E_0}{a}} \hskip -.56cm
\sum_{\stackrel{N_1,N_2}{(N_1+N_2=N)}} 
 \sum_{{\bf k} \in \mc{P}_{N_1}, \, 
{\bf g} \in \mc{P}_{N_2}} \hskip -.42cm
d^{\,(m_1,m_2|n_1,n_2)}_{\bf k, \bf g} \hskip -.42cm 
\sum_{l_1>\dots>l_s \geq
0} \, \sum_{p_1>\dots>p_t \geq 0}\hskip -.4cm 
q^{2 \sum\limits_{i=1}^s l_i  k_i+ 
2\sum\limits_{j=1}^t p_j g_j +N_2} . 
\label{c35a}
\eeq
It may be noted that, 
the summations over $l_i$'s and $p_j$'s appearing
in the above equation can be performed through 
appropriate change of variables~\cite{BFGR08}. As a result,   
we get a simpler expression
for $Z_N^{(m_1,m_2|n_1,n_2)}(aT)$ in \eq{c35a}  as 
\beq
Z_N^{(m_1,m_2|n_1,n_2)}(aT) =q^{\frac{{E}_0}{a}} \hskip -.55cm
\sum_{\stackrel{N_1,N_2}{(N_1+N_2=N)}}
 \sum_{{\bf k} \in \mc{P}_{N_1}, \, {\bf g} \in \mc{P}_{N_2}} \hskip -.3cm
d^{\,(m_1,m_2|n_1,n_2)}_{\bf k, \bf g} \,
q^{-(N+\kappa_s)} \prod_{i=1}^s
\frac{q^{2\kappa_i}}{1-q^{2\kappa_i}} 
\prod_{j=1}^t\frac{q^{2\zeta_j}}{1-q^{2\zeta_j}} \, , 
\label{c35}
\eeq
with $\kappa_i \equiv \sum_{l=1}^i k_l$ and 
$\zeta_j \equiv \sum_{l=1}^j g_l$ representing     
the partial sums associated with the sets  
${\bf k}$ and ${\bf g}$ respectively. 
Inserting the expressions for $Z_N^{(m_1,m_2|n_1,n_2)}(aT)$
in \eq{c35} and $Z_N(aT)$
in \eq{c22}  to  the relation~\eq{c13},
 we derive the partition functions for the  
$BC_N$ type of PF spin chains with SAPSRO \eq{b14} as 
\bea
\mc{Z}_N^{(m_1, m_2|n_1,n_2)}(q)=
\prod_{l=1}^N (1-q^{2l}) \hskip -.55cm
\sum_{\stackrel{N_1,N_2}{(N_1+N_2=N)}} \hskip -.5cm
&& \sum_{{\bf k} \in \mc{P}_{N_1}, \, {\bf g} \in \mc{P}_{N_2}} 
\hskip -.3cm d^{\,(m_1,m_2|n_1,n_2)}_{\, \bf k, \bf g} \,
q^{-(N+\kappa_s)}  \nn \\
&& \times ~ \prod_{i=1}^s \frac{q^{2\kappa_i}}{1-q^{2\kappa_i}} 
\prod_{j=1}^t \frac{q^{2\zeta_j}}{1-q^{2\zeta_j}} \, ,
\label{c36}
\eea 
where from now on we shall use the variable $q=e^{-1/kT}$ instead of $T$.
Let us now try to write the above partition function
as a polynomial function of $q$, 
which is expected for the case of any spin 
 system with finite number of lattice sites. To this end,  
 we define complementary sets of the two sets    
$\{\kappa_1, \kappa_2, \ldots, \kappa_s\}$ and 
$\{\zeta_1, \zeta_2, \ldots, \zeta_t\}$
as $\{\kappa_1', \kappa_2',
\ldots, \kappa'_{N_1-s}\} \equiv
\{1, 2, \ldots, N_1-1, N_1\} \setminus \{\kappa_1, \kappa_2,
\ldots, \kappa_s\} $  and $
 \{\zeta_1', \zeta_2', \ldots, \zeta'_{N_2-t}\} \equiv
\{1, 2, \ldots, N_2-1, N_2\} \setminus 
\{\zeta_1, \zeta_2, \ldots, \zeta_t\}$, respectively.
Using the elements of the sets 
$\{\kappa_1, \kappa_2, \ldots, \kappa_s\}$ and 
$\{\zeta_1, \zeta_2, \ldots, \zeta_t\}$, along with the 
elements of their complementary sets, 
the partition function in 
\eq{c36} can be explicitly written as a polynomial in $q$ as 
\bea
\mc{Z}_N^{(m_1, m_2|n_1,n_2)}(T)  = \hskip -.4 cm  
\sum_{\stackrel{N_1,N_2}{(N_1+N_2=N)}}
 \sum_{{\bf k} \in \mc{P}_{N_1}, \, {\bf g} \in \mc{P}_{N_2}}
\hskip -.3 cm 
d^{\,(m_1,m_2|n_1,n_2)}_{\bf k, \bf g} \hskip -.7cm
&  \qbinom N{N_1}{q^2} \, q^{N_2+2\sum\limits_{i=1}^{s-1}
\kappa_i+2\sum\limits_{j=1}^{t-1} \zeta_j}     \nn \\
\times & \prod\limits_{i=1}^{N_1-s}(1-q^{2\kappa'_i})
\prod\limits_{j=1}^{N_2-t}(1-q^{2\zeta'_j}) \, . 
\label{c38}
\eea
In the above expression, $\qbinom N{N_1}{q^2}$ denotes a 
$q$-binomial coefficient given by  
\beq
\qbinom N{N_1}{q^2}= \frac{\prod\limits_{l=1}^N
(1-q^{2l})}{\prod\limits_{ i=1}^{N_1}(1-q^{2i})\prod\limits_{
j=1}^{N-N_1}(1-q^{2j})} \, ,  \nn
\eeq
which can be expressed as an even polynomial of degree 
$2N_1(N-N_1)$ in $q$ \cite{Ci79}.

\noi \section{Connection with $A_K$ type of supersymmetric PF chains}
\renewcommand{\theequation}{4.{\arabic{equation}}}
\setcounter{equation}{0}
\medskip

In the following, our aim is to establish a connection
between the partition function \eq{c38} and the  
partition functions of some supersymmetric PF spin chains of type $A$. 
To this end, we note that the Hamiltonian of the $A_{N-1}$ type of
$su(m|n)$ supersymmetric PF spin chain is given by~\cite{BUW99,HB00}
\beq
\mc{H}_\mr{PF}^{(m|n)}=\sum_{1\leqslant i<j \leqslant N}
\frac{1-P_{ij}^{(m|n)}}{(\rho_i-\rho_j)^2}  \, . 
\label{d4}
\eeq
It is evident that, for the special case $n=0$,
the above Hamiltonian 
reduces to $ \mc{H}_\mr{PF}^{(m)}$ in \eq{a1} with $\ep=1$.
Moreover, by putting $n=0$ after interchanging 
$m$ and $n$ in \eq{d4}, one gets $\mc{H}_\mr{PF}^{(m)}$ with $\ep=-1$.
There exists a few different but 
equivalent expressions for the partition function of the 
 $su(m|n)$ supersymmetric spin chain \eq{d4} 
 in the literature~\cite{BUW99,HB00,BBH10,BBHS07}. One such expression for 
 the partition function of the spin chain \eq{d4} is given by~\cite{BBHS07}  
\beq
\mathcal{Z}^{(m|n)}_{(A)\, N}(q)= \sum_{ {\bf f} \in 
\mc{P}_{N}} d^{(m|n)}({\bf f}) \, q^{ \sum_{j=1}^{r-1} \mc{F}_j} 
\prod_{j=1}^{N-r}(1- q^{\mc{F}_j'}) \, .
 \label{d3}
\eeq
where $\mbf{f} \equiv \{f_1, f_2 \cdots f_r\}$, 
the partial sums are given by $\mc{F}_j = \sum_{i=1}^j f_i$,  
and the complementary partial sums are defined as
$\{ \mc{F}_1', \mc{F}_1' , \cdots ,  \mc{F}_{N-r}' \}
\equiv \{1, 2, \cdots, N\}-\{\mc{F}_1, \mc{F}_2,
\cdots, \mc{F}_r\}$. Moreover, $d^{(m|n)}({\bf f})$ in the above 
expression is defined through $d_{\, x,y}(\nu)$ in \eq{c33a} as 
\beq
d^{(m|n)}({\bf f})= \prod_{i=1}^r d_{\, m,n}(f_i) \, . 
\label{d3a}
\eeq
Using \Eq{d3a}, one can express the  
spin degeneracy factor
$d^{\, m_1,m_2}_{\, \bf k, \bf g}$ in \eq{c33} as
\beq
d^{\,(m_1,m_2|n_1,n_2)}_{\, \bf k, \bf g}
=d^{(m_1|n_1)}({\bf k}) \, d^{(m_2|n_2)}({\bf g}) \, .
\nn
\eeq
Substituting this factorised form of $d^{\,(m_1,m_2|n_1,n_2)}_{\, \bf k, \bf g}$ 
to \Eq{c38}, we obtain
\bea
&&\mc{Z}_N^{(m_1, m_2|n_1,n_2)}(q)
= \sum_{\stackrel{N_1,N_2}{(N_1+N_2=N)}}
q^{N_2} \qbinom N{N_1}{q^2} \left(  \sum_{ {\bf k} \in 
\mc{P}_{N_1}} d^{(m_1|n_1)}({\bf k})\, q^{2 \sum_{j=1}^{s-1} \kappa_j}
\prod_{j=1}^{N_1-s}(1- q^{2 \kappa_j'}) \right)\nn \\
&&~~~~~~~~~~~~~~~~~~~~~~~~~~~~~~~~~~~~~~~~
\times \left(  \sum_{ {\bf g} \in \mc{P}_{N_2}} \,   d^{(m_2|n_2)}({\bf
g}) \, q^{2 \sum_{j=1}^{t-1} \zeta_j} 
\prod_{j=1}^{N_2-t}(1- q^{2 \zeta_j'}) \right).
\label{d2}
\eea
Using the expression of $\mathcal{Z}^{(m|n)}_{(A)\, N}(q)$
in \eq{d3} for all nontrivial 
cases where $N\geqslant 1$ and $m+n\geqslant 1$, 
and also assuming that $\mc{Z}^{(m|n)}_{(A)\, 0} (q)=1$
and $\mc{Z}^{(0|0)}_{(A)\, N} (q)=\delta_{N,0}$,
we finally rewrite
$\mc{Z}_N^{(m_1, m_2|n_1,n_2)}(q)$ in \eq{d2} as
\beq
\mc{Z}_N^{(m_1, m_2|n_1,n_2)}(q)
= \sum_{N_1=0}^N q^{N-N_1} \qbinom N{N_1}{q^2} 
\mc{Z}^{(m_1|n_1)}_{(A)\, N_1} (q^2)\,
\mc{Z}^{(m_2|n_2)}_{(A)\, N-N_1} (q^2) \, .
\label{d6}
\eeq
Thus we find that the partition function of the 
$BC_N$ type of PF spin chain with SAPSRO \eq{b14} can be expressed
in an elegant way through the partition functions of several
$A_{K}$ type of supersymmetric PF spin chains, where $K\leq N-1$.

We have previously mentioned that,
for a particular choice
of the discrete parameters given by
$m_1=m, ~m_2=0,~n_1=n, ~n_2=0$,
$\mathcal{H}^{(m_1,m_2|n_1,n_2)}$ in  
\eq{b14} reduces to $\mathcal{H}^{(m,0|n,0)}$ in \eq{b16}.
Applying \Eq{d6} for this particular choice
of the discrete parameters and also using  
$\mc{Z}^{(0|0)}_{(A)\, N-N_1} (q^2)=\delta_{N_1,N}$,
we obtain
\bea
\mc{Z}_N^{(m, 0|n,0)}(q) \hskip -.4 true cm
&&= \sum_{N_1=0}^N q^{N-N_1} \qbinom N{N_1}{q^2} 
\mc{Z}^{(m|n)}_{(A)\, N_1} (q^2)\,
\mc{Z}^{(0|0)}_{(A)\, N-N_1} (q^2) \, \nn \\
&&=\mc{Z}^{(m|n)}_{(A)\, N} (q^2)\, .
\label{d7}
\eea
Hence, replacing $q$ by $q^2$ in the RHS of \eq{d3}, 
it is possible to get an explicit expression
for the partition function of 
$\mathcal{H}^{(m,0|n,0)}$ in \eq{b16}.
Since $\mc{Z}_{(A)\, N}^{(m|n)}(q)$ in \eq{d3} 
can be expressed as a polynomial function of $q$, 
\Eq{d7} also implies that the spectrum of
$\mathcal{H}^{(m,0|n,0)}$ would coincide with that 
of the following Hamiltonian $\wt{\mc{H}}_\mr{PF}^{(m|n)}$, 
which is obtained by multiplying 
$\mc{H}_\mr{PF}^{(m|n)}$ in \eq{d4} by a factor of two:
 \beq
\wt{\mc{H}}_\mr{PF}^{(m|n)} =\sum_{1\leqslant i \neq j \leqslant N}
\frac{1-P_{ij}^{(m|n)}}{(\rho_i-\rho_j)^2}  \, . 
\label{scal}
\eeq
As shown in Ref.~\cite{HB00}, the spectrum of such 
$su(m|n)$ supersymmetric 
PF spin chain can be expressed through 
Haldane's motifs which characterize
the irreducible representations of the $Y(gl(m|n)$ Yangian quantum group. 
The motif $\mbs{\delta}$ for the spin chain \eq{scal} is given by a 
$(N-1)$ sequence of $0$'s and $1$'s, i.e. 
$ \mbs{\delta}=(\de_1,\de_2, \cdots , \de_{N-1})$,
with $\de_i\in \{0,1\}$. 
In the non-supersymmetric case where the value of $n$ is taken as zero, 
the motifs of the spin chain \eq{scal} 
obey a `selection 
rule' which forbids the appearance of $m$ number of consecutive $1$'s.
On the other hand,  
$\de_i$'s can freely take the values  
$0$ or $1$ for supersymmetric spin chains
with $m\geqslant 1$ and $n\geqslant 1$. Consequently, it is possible to 
construct $2^{N-1}$ number 
of distinct motifs in the case of supersymmetric
spin chains. All energy levels of the 
spin chain \eq{scal}, in the supersymmetric as well as non-supersymmetric cases, 
can be expressed through the corresponding motifs as~\cite{HB00} 
\beq
E_{\mbs{\de}}= 2\sum_{i=1}^{N-1} j \de_j \, .
\label{motif}
\eeq
Hence, due to \Eq{d7},
it follows that the spectrum of $\mathcal{H}^{(m,0|n,0)}$ in \eq{b16}
is also be given by $E_{\mbs{\de}}$ in the above equation. 
In particular, for the supersymmetric case,
the motif $ \mbs{\delta}=(0,0, \cdots ,0)$ gives 
the ground state energy of this Hamiltonian  
 as $\mc{E}^{(m,0|n,0)}_{min} =0$
and the motif $ \mbs{\delta}=(1,1, \cdots ,1)$ gives the corresponding highest 
state energy as $\mc{E}^{(m,0|n,0)}_{max} =N^2 -N$. 
The degeneracy of each energy level in \eq{motif} can also be computed
for all possible values of $m$ and $n$,
by taking appropriate limits of the supersymmetric Schur polynomials~\cite{HB00}.
Thus it is possible to find out the full 
spectrum of the supersymmetric spin chain  \eq{b16}, by
using our key result that this spectrum coincides with 
that of the $A_{N-1}$ type of $su(m|n)$ supersymmetric PF spin chain
\eq{scal}.

We have already mentioned that, the lattice sites 
of $\mathcal{H}^{(m,0|n,0)}$ in \eq{b16} and 
$\wt{\mc{H}}_\mr{PF}^{(m|n)}$ in \eq{scal}
are determined through the zero points of the 
generalized Laguerre polynomial $L_N^{\beta -1}$ and 
the zero points of the Hermite polynomial $H_N$ respectively.
Thus the lattice sites of these two Hamiltonians 
are quite different in nature.
However, since $\mathcal{H}^{(m,0|n,0)}$ 
and $\wt{\mc{H}}_\mr{PF}^{(m|n)}$
share exactly same spectrum, these two Hamiltonians must be related
through a unitary transformation like
\beq
\mathcal{H}^{(m,0|n,0)} \, = \, 
\mc{S}^{(m|n)}_\beta \, \wt{\mc{H}}_\mr{PF}^{(m|n)}
\left(\mc{S}^{(m|n)}_\beta \right)^\dagger \, .
\label{sim}
\eeq
Even though we do not know the explicit form 
of $\mc{S}^{(m|n)}_\beta$, it is possible
to find out the asymptotic form of this 
operator at $\beta \to \infty$ limit by using the following conjecture. 
For any $N\geq 2$, let us order the zero points of the 
of the Hermite
polynomial $H_N$ and the generalized Laguerre polynomial $L_N^{\beta -1}$
on the real line as  
$\rho_1 > \rho_2> \cdots > \rho_N$ and $y_1>y_2>\cdots >y_N$ respectively. 
Then, based on numerical results, it has been conjectured that
these zero points would satisfy the asymptotic relations
given by~\cite{BBB14}
\beq
 \underset{\beta \rightarrow \infty    }{\mr{lim}}   \,
 \frac{y_i+y_j}{(y_i-y_j)^2} = \frac{1}{(\rho_i -\rho_j)^2} \, , 
 \label{con}
\eeq
where $1\leq i<j\leq N$. Using this conjecture, 
it is easy to see that the $\beta \to \infty$ limit of
 $\mathcal{H}^{(m,0|n,0)}$ in \eq{b16} yields 
$\wt{\mc{H}}_\mr{PF}^{(m|n)}$ in \eq{scal}.
Hence \Eq{sim} would be satisfied in this limit if 
we take the asymptotic form of $\mc{S}^{(m|n)}_\beta$ as 
$\lim_{\beta \to \infty} \mc{S}^{(m|n)}_\beta= \one$.

\noi \section{Extended boson-fermion duality for $BC_N$ type of
PF chains with SAPSRO }
\renewcommand{\theequation}{5.{\arabic{equation}}}
\setcounter{equation}{0}
\medskip
Boson-fermion duality relations involving the partition functions 
of various $A_{N-1}$ type of supersymmetric spin chains 
with long-range interaction have been established in the literature 
\cite{BUW99,HB00,BB06, BBHS07}. 
Subsequently, a similar type of duality relation has been studied  
for the case of $BC_N$ type of PF spin chains associated with 
the supersymmetric analogue of spin reversal operators~\cite{BFGR09}.
More precisely, it has been found in the latter reference that  
\beq
\mc Z_{\epsilon,\epsilon^{\prime}}^{(m|n)}(q)
=q^{N^2}\mc Z_{-\epsilon^{\prime},-\epsilon}^{(n|m)}(q^{-1}) \, ,
\label{g17}
\eeq
where $\mc Z_{\epsilon,\epsilon^{\prime}}^{(m|n)}(q)$ 
represents the partition function for the Hamiltonian 
$\mc {H}^{(m|n)}_{\ep,\ep'}$ in \eq{b11a}. 
It is evident that the duality relation \eq{g17} not only  
involves the exchange of bosonic and fermionic 
degrees freedom, but also the exchange of the two
discrete parameters $\ep$ and $\ep'$ along with their sign change. 
For the purpose of gaining some deeper understanding for such change 
of the two discrete parameters, in the following we aim to study 
the duality relation for the case of $BC_N$ type of PF chains \eq{b14}
associated with SAPSRO. 

To begin with, we define the star operator $\mc{S}$:
$\mbf{\Sigma}^{(m_1,m_2|n_1,n_2)}\rightarrow 
\mbf{\Sigma}^{(m_1,m_2|n_1,n_2)}$ as
\beq
 \mc{S} \ket {s_1,s_2,\cdots,s_N}
=(-1)^{\sum\limits_{j=1}^N j\pi(s_j)}\ket{s_1,s_2,\cdots,s_N} \, .
\label{g1}
\eeq
It is easy to verify that $\mc{S}$ operator is self-adjoint and 
$\mc{S}\circ \mc{S}$ 
is the identity in $\mbf{\Sigma}^{(m_1,m_2|n_1,n_2)}$.
Next, we consider the Hilbert space $\mbs{\Sigma}^{(n_2,n_1|m_2,m_1)}$,
and denote the corresponding supersymmetric spin exchange operator
and the SAPSRO as $P_{ij}^{(n|m)}$ and 
$P_i^{(n_2,n_1|m_2,m_1)}$ respectively. The Hamiltonian 
$\mc H^{(n_2,n_1|m_2,m_1)}$ associated with this Hilbert 
space is evidently obtained from 
$\mc H^{(m_1,m_2|n_1,n_2)}$ in \eq{b14} through the 
replacements: $m_1\to n_2$, $m_2\to n_1$, $n_1\to m_2$ and  
$n_2\to m_1$. In analogy with the   
basis vectors of $\mbs{\Sigma}^{(m_1,m_2|n_1,n_2)}$ and the 
ranges of the corresponding spin components in \eq{b12},
we assume that $\mbs{\Sigma}^{(n_2,n_1|m_2,m_1)}$ is spanned 
by orthonormal state vectors like  
$\ket{\bar{s}_1,\cdots,\bar{s}_N}$, where the components of 
$\bar{s}_i \equiv (\bar{s}_i^1,
\bar{s}_i^2,\bar{s}_i^3)$ are 
taking values within the ranges 
\begin{subequations}
\bea
&&\bar{s}_i^1\equiv \pi(\bar{s}_i) = \left \{  
\begin{array}{ll} 
0,  & \mbox{~for bosons, } \\
1, &   \mbox{~for fermions, } 
\end{array} 
\right.
\label{g2a}\\
&&\bar{s}_i^2\equiv f(\bar{s}_i) = \left \{  
\begin{array}{ll} 
0,  & \mbox{~for positive parity under SAPSRO,} \\
1, &   \mbox{~for negative parity under SAPSRO, } 
\end{array} 
\right.
\label{g2b}\\
&&\bar{s}_i^3 \in \left \{ \hskip -.47 cm  
\begin{array}{llll} 
&\mbox{~ $\{1,2, \cdots,n_2 \},
~\mr{if}~\pi(s_i)=0~ \mr{and} ~f(s_i)=0$, } \\
&~\mbox{~$\{1, 2, \cdots, n_1 \},
~\mr{if}~\pi(s_i)=0 ~ \mr{and} ~ f(s_i)=1 $, } \\
&~\mbox{~$\{1, 2, \cdots, m_2 \},
~\mr{if}~\pi(s_i)=1~ \mr{and} ~f(s_i)=0 $, } \\
&~\mbox{~$\{1, 2, \cdots, m_1\},
~\mr{if}~\pi(s_i)=1~ \mr{and} ~f(s_i)=1 $. } 
\end{array} 
\right.
\label{g2c}
\eea
\label{g2}
\end{subequations}
It is evident that the  spaces 
$\mbf{\Sigma}^{(m_1,m_2|n_1,n_2)}$ and 
$\mbs{\Sigma}^{(n_2,n_1|m_2,m_1)}$ have the same
dimension given by $(m+n)^N$. 
Let us now define an invertible operator ${\chi}^{(m_1,m_2|n_1,n_2)}$:
$\mbf{\Sigma}^{(m_1,m_2|n_1,n_2)}
\rightarrow \mbf{\Sigma}^{(n_2,n_1|m_2,m_1)}$
by
\beq
\chi^{(m_1,m_2|n_1,n_2)}\ket {s_1,s_2,\cdots,s_N}
=\ket{\bar{s}_1,\bar{s}_2,\cdots,\bar{s}_N} \, ,
\label{g3}
\eeq
where
\[
\bar{s}_i^1=1-s_i^1 ,~ \bar{s}_i^2=1-s_i^2, ~\bar{s}_i^3=s_i^{3}.
\]
From the above relation it is clear that, 
if $s_i$ represents a bosonic  
(fermionic) spin with parity $\pm1$ under SAPSRO, 
then $\bar{s}_i$ would represent a fermionic (bosonic) spin 
with parity $\mp1$ under SAPSRO.
Using \Eq{g3}, it is easy to check that 
${\chi^{(m_1,m_2|n_1,n_2)}}^\dagger =\chi^{(n_2,n_1|m_2,m_1)} $
and $\chi^{(n_2,n_1|m_2,m_1)} \circ \chi^{(m_1,m_2|n_1,n_2)}$ 
is the identity in $\mbf{\Sigma}^{(m_1,m_2|n_1,n_2)}$.
Subsequently, we define the operator 
$\mc U^{(m_1,m_2|n_1,n_2)}$:
$\mbf{\Sigma}^{(m_1,m_2|n_1,n_2)}\rightarrow 
\mbf{\Sigma}^{(n_2,n_1|m_2,m_1)}$ as the composition
\beq
\mc U^{(m_1,m_2|n_1,n_2)}=\chi^{(m_1,m_2|n_1,n_2)}\circ  \mc{S}.
\label{g4}
\eeq
By using the above mentioned properties of 
$\mc{S}$ and $\chi^{(m_1,m_2|n_1,n_2)}$, it is easy to show that 
$\mc U^{(m_1,m_2|n_1,n_2)}$ in \eq{g4} is an unitary operator 
satisfying the relation
\beq
{\mc U^{(m_1,m_2|n_1,n_2)}}^{\dagger}
={\mc U^{(m_1,m_2|n_1,n_2)}}^{-1}=\mc{S}\circ 
\chi^{(n_2,n_1|m_2,m_1)}.
\label{g5}
\eeq
Using Eqs.~\eq{g1} and \eq{g3}, 
and closely following the procedure of Ref.~\cite{BBHS07} 
for establishing boson-fermion duality relation in the case of 
$A_{N-1}$ type of supersymmetric HS spin chain,  
it is straightforward to show that 
$\mc U^{(m_1,m_2|n_1,n_2)}P_{ij}^{(m|n)}
=-P_{ij}^{(n|m)}\mc U^{(m_1,m_2|n_1,n_2)}$, 
or equivalently
\beq
{\mc U^{(m_1,m_2|n_1,n_2)}}^{-1}P_{ij}^{(n|m)}
\mc U^{(m_1,m_2|n_1,n_2)}
=-P_{ij}^{(m|n)}.
\label{g6}
\eeq
Next, by using Eqs.~\eq{b13}, \eq{g1}, \eq{g3} and \eq{g4}, 
we find that 
\beq
\mc U^{(m_1,m_2|n_1,n_2)}P_i^{(m_1,m_2|n_1,n_2)}\ket{s_1,\cdots,s_N}
=(-1)^{f(s_i)}(-1)^{\sum\limits_{j=1}^N j\pi(s_j)}
\ket{\bar{s_1},\cdots,\bar{s_N}} \, , 
\label{g7}
\eeq
and
\beq
P_i^{(n_2,n_1|m_2,m_1)}
\mc U^{(m_1,m_2|n_1,n_2)}\ket{s_1,\cdots,s_N}
=(-1)^{f(\bar{s}_i)}(-1)^{\sum\limits_{j=1}^N j\pi(s_j)}
\ket{\bar{s_1},\cdots,\bar{s_N}} \, . 
\label{g8}
\eeq
Since, due to Eqs.~\eq{g3}, it follows that 
$(-1)^{f(s_i)}=-(-1)^{f(\bar{s_i})} $,  
comparing Eq.~\eq{g7} with Eq.~\eq{g8} we find that 
\[
\mc U^{(m_1,m_2|n_1,n_2)}P_i^{(m_1,m_2|n_1,n_2)}
=-P_i^{(n_2,n_1|m_2,m_1)}\mc U^{(m_1,m_2|n_1,n_2)} \, ,
\]
or, equivalently
\beq
{\mc U^{(m_1,m_2|n_1,n_2)}}^{-1}P_i^{(n_2,n_1|m_2,m_1)}
\mc U^{(m_1,m_2|n_1,n_2)}=-P_i^{(m_1,m_2|n_1,n_2)}.
\label{g11}
\eeq
With the help of 
Eqs.~\eq{b14}, \eq{g6} and \eq{g11}, we obtain
\begin{eqnarray}
&&\mc H^{(m_1,m_2|n_1,n_2)}+
{\mc U^{(m_1,m_2|n_1,n_2)}}^{-1}\mc H^{(n_2,n_1|m_2,m_1)}
\mc U^{(m_1,m_2|n_1,n_2)}\nonumber\\
&&\hskip 2 cm =2\sum\limits_{i\neq j}
\left[(\xi_i-\xi_j)^{-2}+(\xi_i+\xi_j)^{-2}\right]
+2\beta\sum\limits_i\xi_i^{-2}
=N^2,
\label{g12}
\end{eqnarray}
where the last sum has been  
derived in Ref.~\cite{BFGR08}. Since the Hamiltonians 
$\mc H^{(n_2,n_1|m_2,m_1)}$ and 
${\mc U^{(m_1,m_2|n_1,n_2)}}^{-1}\mc H^{(n_2,n_1|m_2,m_1)}
\mc U^{(m_1,m_2|n_1,n_2)}$ 
are isospectral, Eq.~\eq{g12} implies that 
the spectra of $\mc H^{(m_1,m_2|n_1,n_2)}$ 
and $\mc H^{(n_2,n_1|m_2,m_1)}$ are `dual' to each other. 
More precisely, the eigenvalues of $\mc H^{(m_1,m_2|n_1,n_2)}$ 
and $\mc H^{(n_2,n_1|m_2,m_1)}$  
are related as 
\beq
 \mc E^{(m_1,m_2|n_1,n_2)}_i
  =N^2 -  \mc E^{(n_2,n_1|m_2,m_1)}_i \, . 
\label{g13a}
  \eeq
Using the above equation, we obtain a novel type of 
duality relation between the 
partition functions of $\mc H^{(m_1,m_2|n_1,n_2)}$ 
and $\mc H^{(n_2,n_1|m_2,m_1)}$ as  
\beq
\mc Z^{(m_1,m_2|n_1,n_2)}(q)
=q^{N^2}\mc Z ^{(n_2,n_1|m_2,m_1)}(q^{-1}).
\label{g13}
\eeq
It is interesting to observe that this duality relation   
not only involves the exchange of bosonic and 
fermionic degrees of freedom, but also involves the exchange of           
positive and negative parity degrees of freedom associated with 
SAPSRO. Therefore, the duality
relation \eq{g13} can be interpreted as a nontrivial extension 
of the usual boson-fermion duality relation which holds  
for the case of $A_{N-1}$ type of supersymmetric spin chains. 
It is also interesting to note that, applying the relation 
\eq{g13a} in the special case where 
$n_1=m_2$ and $n_2=m_1$,  
the spectrum of the Hamiltonian ${\mc H}^{(m_1,m_2|m_2,m_1)}$
can be shown to be invariant under $ \mc{E} \mapsto N^2-\mc{E}$, i.e.,
to be symmetric about the mean energy $N^2/2$. 

We have mentioned in Sec.\,2 that, for the special values of  
discrete parameters appearing in \eq{b15},
it is possible to construct 
an unitary transformation which  
maps $P_i^{(m_1,m_2|n_1,n_2)}$
to  $P_i^{\ep,\ep^{\prime}}$ and keeps 
$P_{ij}^{(m|n)}$ invariant. 
It is interesting to observe that \Eq{b15}
remains invariant under the simultaneous transformations
given by:   $m_1\to n_2$, $m_2\to n_1$, $n_1\to m_2$, 
$n_2\to m_1$ and 
$\ep \to -\ep'$, $\ep' \to -\ep$. Hence,  
it is also possible to construct an unitary transformation which  
would map $P_i^{(n_2,n_1|m_2,m_1)}$
to  $P_i^{-\ep',-\ep}$ and keep $P_{ij}^{(n|m)}$ invariant.
Due to the existence of such unitary transformations in the special case \eq{b15},
$\mc H^{(m_1,m_2|n_1,n_2)}$ in \eq{b14}
and related $\mc H^{(n_2,n_1|m_2,m_1)}$ become equivalent 
to the Hamiltonians $\mc {H}^{(m|n)}_{\ep,\ep'}$ in \eq{b11a}
and related  $\mc {H}^{(n|m)}_{-\ep',-\ep}$ respectively.
Consequently, for the special values of discrete parameters
given in \eq{b15}, our duality relation \eq{g13} would 
naturally reproduce the previously obtained 
duality transformation \eq{g17}. 

Next, let us now investigate whether 
extended boson-fermion duality relation like  
\eq{g13} holds for some other quantum spin chains 
associated with SAPSRO. To this end, we consider 
a class of one dimensional spin chains 
with Hamiltonian of the form 
\beq
\hat{\mathcal{H}}^{(m_1,m_2|n_1,n_2)} \! = \!
\sum\limits_{i\neq j}\left[w_{ij}(1-P_{ij}^{(m|n)})+
\tilde{w}_{ij}(1-\widetilde{P}_{ij}^{(m_1,m_2|n_1,n_2)})\right]
+\sum\limits_i \! w_i\left(1-P_i^{(m_1,m_2|n_1,n_2)}\right),
\label{g14}
\eeq
where $w_{ij}$, $\tilde{w}_{ij}$, $w_i$ are arbitrary  
real parameters. Clearly, the above Hamiltonian would represent a 
non-integrable system for almost all values of these parameters.
Using again Eqs.~\eq{g6} and \eq{g11}, we find that 
\beq
\hat{\mathcal{H}}^{(m_1,m_2|n_1,n_2)}+
{\mc U^{(m_1,m_2|n_1,n_2)}}^{-1}
\hat{\mathcal{H}}^{(n_2,n_1|m_2,m_1)}\mc U^{(m_1,m_2|n_1,n_2)}
=W ,
\label{g15}
\eeq
where $W=2(\sum_{i\neq j}(w_{ij}
+\tilde{w}_{ij})+\sum_i w_i)$.
Using this relation and proceeding as before, we obtain a duality relation
given by 
\beq
\hat{\mc Z }^{(m_1,m_2|n_1,n_2)}(q)
=q^{W}\hat{\mc Z}^{(n_2,n_1|m_2,m_1)}(q^{-1}) \, ,
\label{g16}
\eeq
where $\hat{\mc Z }^{(m_1,m_2|n_1,n_2)}(q)$ denotes 
the partition function of 
$\hat{\mc H}^{(m_1,m_2|n_1,n_2)}$.
Hence, the extended boson-fermion duality relation  
can be applied to a wide range of spin chains of the form \eq{g14}. 
In the following, however, we shall restrict its application only 
for the case of $BC_N$ type of PF chains \eq{b14}
associated with SAPSRO. Indeed, in the next section, at first   
we shall compute the ground state energies for the spin 
chains \eq{b14} with the help of the freezing trick and subsequently
derive the corresponding highest state energies by using this    
duality relation.  

\bigskip

\noi \section{Ground state and highest state energies 
for PF chains with SAPSRO}
\renewcommand{\theequation}{6.{\arabic{equation}}}
\setcounter{equation}{0}
\medskip
It is well known that the spectra of the  
$A_{N-1}$ type of PF spin chain \eq{a1} and its supersymmetric
generalization \eq{d4} are equispaced within
the corresponding lowest and highest energy levels. This result 
follows from the fact that corresponding partition functions  
can be expressed as some polynomials in
$q$, where all consecutive powers of $q$
(within the allowed range) appear with 
positive integer coefficients. 
It has been shown in Ref.~\cite{BBB14} that spectrum 
for the $BC_N$ type of PF chains \eq{b14} are also equispaced  
in the special case where either  
bosonic or fermionic spins are present. 
Using the expression of the partition function \eq{d6} and following
the arguments of the later reference, it can be shown that 
the spectra for the $BC_N$ type of PF chains \eq{b14} 
are also equispaced when both of the  
bosonic and fermionic spins are present, i.e., when $m,n\geqslant 1$.
At present, our aim is to compute the lower and the upper limits 
of such equispaced spaced spectra, i.e., 
the ground state and the highest state energies 
of the Hamiltonian ${\mc H}^{(m_1,m_2|n_1,n_2)}$ in \eq{b14} 
for the cases where $m,n \geqslant 1$.

In Sec.~4 it has been shown that, for the particular choice 
of the discrete parameters given by
$m_1=m, ~m_2=0,~n_1=n, ~n_2=0$,
the spectrum of the Hamiltonian ${\mc H}^{(m_1,m_2|n_1,n_2)}$
coincides with that of 
$\wt{\mc{H}}_\mr{PF}^{(m|n)}$ in \eq{scal}.
By using such coincidence, we have found the 
ground state and the highest state energies 
of the Hamiltonian ${\mc H}^{(m,0|n,0)}$
 as $\mc{E}^{(m,0|n,0)}_{min} =0$ and
 $\mc{E}^{(m,0|n,0)}_{max} =N^2 -N$, respectively.
The above mentioned method of calculating the ground state 
and the highest state energies is clearly not applicable for 
more general cases where $m_2$ or $n_2$ takes nontrivial value. 
However, by using the freezing trick, it is possible to compute 
the ground state energy 
of ${\mc H}^{(m_1,m_2|n_1,n_2)}$ in \eq{b14}
for all cases where $m,n \geqslant 1$.
To this end, 
we consider \Eq{c12} which implies that 
\beq
\mathcal{E}_{min}^{(m_1,m_2|n_1,n_2)}=
\lim_{a \rightarrow \infty}
\frac{1}{a}(E_{min}^{(m_1,m_2|n_1,n_2)}-E_0),
\label{e1}
\eeq
where $E_0$ is the known ground state energy \eq{E0} 
of the $BC_N$ type of scalar Calogero model and 
$E_{min}^{(m_1,m_2|n_1,n_2)}$ represents the ground state  
energy of the $BC_N$ type of
spin Calogero model \eq{c1}.  
Using \Eq{c32}, we can express the latter ground state energy
as $E_{min}^{(m_1,m_2|n_1,n_2)}= a|{\bf{r}}|_{min}+E_0$, 
where $|{\bf{r}}|_{min}$ denotes the minimum value of $|{\bf{r}}|$
 for all possible choice of the multi-index ${\bf r}$
compatible with the conditions $1)-3)$ of Sec.~$3$.  
Substituting this expression of $E_{min}^{(m_1,m_2|n_1,n_2)}$ 
in \Eq{e1}, we find that the ground state energy of the 
spin chain \eq{b14} is given by
\beq
\mathcal{E}_{min}^{(m_1,m_2|n_1,n_2)}=|{\bf{r}}|_{min} \,.
\label{e2}
\eeq
For the purpose of finding out the explicit value of
$\mathcal{E}_{min}^{(m_1,m_2|n_1,n_2)}$, 
in the following we divide the spin chains \eq{b14} 
with $m, n \geqslant 1$ into two distinct classes. \\
\vskip .2 cm 

\noi {\bf Case~I}: Here, we consider all spin chains \eq{b14} with  
$m_1 \geqslant 1$ and $n\geqslant 1$. In this case, there exists 
at least one type of bosonic spin with positive parity (under SAPSRO).
From the conditions $2)$ and $3)$ of Sec.~$3$ it follows that,
all $s_i$'s can be filled up by this type of spin
if we choose the 
corresponding $\mbf{r}$ as $(0,0,\cdots ,0)$.
So, using \eq{e2} we obtain
\beq
\mathcal{E}_{min}^{(m_1,m_2|n_1,n_2)}=0.
\label{e3}
\eeq
\hskip .01cm {\bf Case~II}: 
Let us consider all spin chains \eq{b14} with  
$m_1 =0$, $m_2\geqslant 1$ and $n\geqslant 1$.
In this case, there exist
$m_2$ types of bosonic spins with negative parity.
Furthermore, if $n_1 >0$, there exist $n_1$ types of 
fermionic spins with positive parity.   
Due to the condition $2)$ of Sec.~$3$,
$s_i$'s can be filled up by only these $n_1$ types of spin states 
corresponding to $r_i=0$. Since these are fermionic spin states, 
due to the condition $3)$ of Sec.~$3$,
at most $n_1$ number of consecutive $r_i$'s 
are allowed to take the zero value. 
Now if $N\leqslant n_1 $, then it is evident that
$\mathcal{E}_{min}=0$. For $N>n_1$, we can take $r_i=1$ for 
the remaining $N-n_1$ number of positions, 
and fill up the corresponding $s_i$'s 
by any of the $m_2$ types of bosonic spins
with negative parity. Consequently, we find that 
the configuration   
\beq
{\bf r}=(\overbrace{0, \ldots, 0}^{n_1},
\,\overbrace{1, \ldots, 1}^{N-n_1})\, \nn
\eeq
yields $|{\bf{r}}|_{min}$ in \Eq{e2}.
Thus for all possible spin chains with $m_1=0$ and $n\geqslant 1$, 
we obtain 
\beq
\mathcal{E}_{min}^{(m_1,m_2|n_1,n_2)}=max\,\{N-n_1, 0\}.
\label{e4}
\eeq

\medskip

It is interesting to observe that the highest
eigenvalue of $\mc H^{(m_1,m_2|n_1,n_2)}$ can be determined
in terms of the lowest eigenvalue of $\mc H^{(n_2,n_1|m_2,m_1)}$  
by using the duality relation \eq{g13a}. Hence, 
for the purpose of computing the highest energy eigenvalues of 
the spin chains \eq{b14} for $m, n \geqslant 1$, it is convenient to divide 
these spin chains into following two distinct classes. 
At first, we consider all spin chains \eq{b14} 
with $n_1\geqslant 1$, $n_2 =0$ and $m\geqslant 1$.
With the help of Eqs.~\eq{g13a} and \eq{e4}, 
we find that the highest energy eigenvalues 
for this class of spin chains are given by 
\beq
\mathcal{E}_{max}^{(m_1,m_2|n_1,n_2)}=N^2-max\,\{N-m_2, 0\}.
\label{e5}
\eeq
Finally, we consider all spin chains \eq{b14} with
$n_2 \geqslant 1$ and $m\geqslant 1$.
Using Eqs.~\eq{g13a} and \eq{e3}, we obtain the highest energy eigenvalues 
for this class of spin chains as 
\beq
\mathcal{E}_{max}^{(m_1,m_2|n_1,n_2)}=N^2.
\label{e6}
\eeq

\bigskip

\noi \section{Some spectral properties of 
PF spin chains with SAPSRO}
\renewcommand{\theequation}{7.{\arabic{equation}}}
\setcounter{equation}{0}
\medskip
It may be noted that, 
with the help of symbolic software package like Mathematica,
the partition function 
$\mc{Z}_N^{(m_1, m_2|n_1,n_2)}(q)$ in 
\eq{d6} can be explicitly written as a polynomial of $q$
for a wide range of values of the parameters 
$m_1$, $m_2$, $n_1$, $n_2$, and $N$. If 
 the term $q^{\mc{E}_i}$ appears in such a polynomial
 with (positive) integer valued coefficient $c({\mc{E}_i})$, 
then $\mc{E}_i$ would represent an energy level  
 with degeneracy factor or `level density'
$c({\mc{E}_i})$ in the corresponding spectrum.
Since the sum of these degeneracy factors
for the full spectrum is given by the dimension
of the corresponding Hilbert space,     
the normalized level density $d(\mathcal{E}_i)$ is obtained through
the relation $d(\mathcal{E}_i)= c(\mathcal{E}_i)/(m+n)^N$.
In this way, it is possible to compute 
the level density distribution for the  
$BC_N$ type of PF chains with SAPSRO. 
By using such procedure, 
it has been found earlier that 
the  level densities of both 
$A_{N-1}$ type of PF spin chain \eq{a1} and its supersymmetric 
extension \eq{d4} follow the Gaussian distribution with high
degree of accuracy for sufficiently large number of 
lattice sites~\cite{BFGR08epl,BB09}.
Furthermore,  the level densities of 
the $BC_N$ type of PF chain with usual spin reversal operator 
 and its extension on a superspace \eq{b11a}
have been found to satisfy the Gaussian distribution for  
sufficiently large values of $N$~\cite{BFGR08,BFGR09}.
The Gaussian behaviour of the level
density distributions at $N\to \infty$ limit
has also been established analytically
for the case of several $A_{N-1}$ type of  
spin chains and related vertex models~\cite{EFG10,BB12}.

In this section, at first we shall study the level density distributions
of the $BC_N$ type of PF spin chains with SAPSRO \eq{b14}
for the case of finite 
but sufficiently large number of lattice sites.
However it has been mentioned earlier that, 
for the special case  \eq{b15}, 
$\mathcal{H}^{(m_1,m_2|n_1,n_2)}$ in \eq{b14}
becomes equivalent to the previously studied Hamiltonian 
$\mathcal{H}^{(m|n)}_{\ep \ep'}$ 
in \eq{b11a}. We have also shown that,
in another special case given by
$m_1=m, ~m_2=0,~n_1=n, ~n_2=0$,
the spectrum of the Hamiltonian ${\mc H}^{(m_1,m_2|n_1,n_2)}$
coincides with that of the $A_{N-1}$ type of 
supersymmetric PF spin chain \eq{scal}.
For the purpose of excluding these two
special cases for which spectral properties are already known,
in the following we shall restrict our attention to the spin chains  
\eq{b14} where $m_1$, $m_2$, $n_1$ and $n_2$ 
are taken as positive integers satisfying the conditions 
$\v m_1-m_2 \v > 1 $ and $\v n_1-n_2 \v > 1 $.  
To begin with, let us compute the  
 mean ($\mu$) and the variance ($\sigma$) for the spectrum  
of the 
Hamiltonian $\mc{H}^{(m_1,m_2|n_1,n_2)}$, which are given by
the relations
\beq 
\mu=\frac{ \tr\left[\mc{H}^{(m_1,m_2|n_1,n_2)}\right] } {(m+n)^N}, ~~~~
\sigma^2= 
\frac{\tr \left[(\mc{H}^{(m_1,m_2|n_1,n_2)})^2\right]}{(m+n)^N}-\mu^2 
\, .
\label{f1}
\eeq
Defining four parameters such as  
$\tau_1 \equiv m_1+m_2+n_1+n_2$, $\tau_2 \equiv m_1-m_2+n_1-n_2$, 
$\tau_3 \equiv m_1+m_2-n_1-n_2$, and 
$\tau_4 \equiv m_1-m_2-n_1+n_2$, 
and applying Eqs.~\eq{b10} as well as \eq{b13}, 
we obtain a set of trace relations given by  
\bea
 &&\tr \left[\one \right]=\tau_1^N, ~\tr\left[ P_i^{(m_1,m_2|n_1,n_2)}
\right]=\tau_2 \, \tau_1^{N-1}, ~\tr \left[P_{ij}\right]=\tr 
\left[\widetilde{P}_{ij}^{(m_1,m_2|n_1,n_2)}\right]
=\tau_3 \, \tau_1^{N-2}, \nn \\
&&\tr\left[P_{ij}P_i^{(m_1,m_2|n_1,n_2)}\right]
=\tr
\left[\widetilde{P}^{(m_1,m_2|n_1,n_2)}_{ij}
P_i^{(m_1,m_2|n_1,n_2)}\right]
=\tau_4 \, \tau_1^{N-2}, \nn \\
&&\tr\left[P_{ij} P_k^{(m_1,m_2|n_1,n_2)}\right]=\tr\left[\widetilde{P} _
{ij}^{(m_1,m_2|n_1,n_2)}
P_k^{(m_1,m_2|n_1,n_2)}\right] 
=\tau_2 \, \tau_3 \, \tau_1^{N-3}, \nn \\
&&\tr\left[P_{ij}P_{jl}\right]
=\tr\left[P_{ij}\widetilde{P}_{jl}^{(m_1,m_2|n_1,n_2)}\right]
=\tr\left[\widetilde{P}_{ij}^{(m_1,m_2|n_1,n_2)}\widetilde{P}_{jl}^{(m_1,m_2|n_1,
n_2)} \right] =\tau_1^{ N-2},
\nn \\ 
&&\tr \left[
P_{ij}P_{kl}\right]=\tr\left[P_{ij}\widetilde{P}_{kl}^
{(m_1,m_2|n_1,n_2)}\right]=
\tr\left[\widetilde{P}_{ij}^{(m_1,m_2|n_1,n_2)}
\widetilde{P}_{kl}^{(m_1,m_2|n_1,
n_2)} \right] = 
\tau_3^2 \, \tau_1^{N-4}, \nn \\
&&\tr\left[P_{ij}\widetilde{P}_{ij}^{(m_1,m_2|n_1,n_2)}\right]
=\tr\left[P_i^{(m_1,m_2|n_1,n_2)}P_j^{(m_1,m_2|n_1,n_2)}\right]
=\tau_2^2 \, \tau_1^{N-2}, \nn
\eea 
where
it is assumed that $i, j, k, l$ are all different indices.
Substituting the explicit form of 
$\mathcal{H}^{(m_1,m_2|n_1,n_2)}$ in \eq{b14} 
to \Eq{f1} and using the aforementioned
trace formulae, we get
\beq
\mu=\left(1-\frac{\tau_3}{\tau_1^2}\right)\sum_{i\neq j}~(h_{ij}
+\widetilde{h}_{ij})+\left(1-\frac{\tau_2}{\tau_1}\right)
\sum_{i=1}^{N}h_i \, ,  
\label{f2}
\eeq
and
\begin{align}
\sigma^2 
&=2 \left(1-\frac{\tau_3^2}{\tau_1^4}\right)\sum_
{i\neq
j}(h^2_{ij}+\widetilde{h}^2_{ij})
+4\left(\frac{\tau_1^2\tau_2^2-\tau_3^2}{\tau_1^4}\right)
\sum_{i\neq j}h_{ij}\widetilde{h}_{ij}  
+\left(1-\frac{\tau_2^2}{\tau_1^2}
\right)\sum_{i=1}^{N}h_i^2 
\nn \\
&+\frac{4(\tau_1 \tau_4-\tau_2 \tau_3)}{\tau_1^3}\,\sum_{i\neq
j}(h_{ij}+\widetilde{h}_{ij})h_i 
+\frac{16\, mn}{\tau_1^4}\, {\sum_{i,j,k}}^{\prime}(h_{ij}
+\widetilde{h}_{ij})(h_{jk}
+\widetilde{h}_{jk}) 
\label{sigma},
\end{align}
where $h_{ij} \equiv 1/(\xi_i-\xi_j)^2$,  
$\widetilde{h}_{ij} \equiv 1/(\xi_i+\xi_j)^2$, 
$h_i \equiv \beta/\xi_i^2$, and 
the symbol ${\sum\limits_{i,j,k}}^{\!\prime}$ 
denotes summation over $i\neq j\neq k\neq i$. 
Using equations~\eq{f2} and \eq{sigma} along with   
the identities given by~\cite{Ah78,AM83,BFGR08}
\bea
&&\sum_{i\neq j}~(h_{ij}+\widetilde{h}_{ij})=\frac{N}{2}(N-1), 
~\sum_{i=1}^{N}h_i=\frac{N}{2},  \nn \\
&&\sum_{i\neq
j}~(h_{ij}^2+\widetilde{h}_{ij}^2)=
\frac{N(N-1)}{72(1+\beta)}\left[2\beta(2N+5)+4N+1\right] ,
\nn \\ 
&&\sum_{i=1}^{N}h_i^2=\frac{N(N+\beta)}{4(1+\beta)},
~\sum_{i\neq j}~h_{ij}\widetilde{h}_{ij}=\frac{N(N-1)}{16(1+\beta)},
~\sum_{i\neq j}(h_{ij}+\widetilde{h}_{ij})h_i=\frac{N}{4}(N-1),
  \nn \\
&&
{\sum_{i,j,k}}^{\prime}(h_{ij}+\widetilde{h}_{ij})(h_{jk}
+\widetilde{h}_{jk})=\frac{2}{9}N(N-1)(N-2) \, ,  
\eea
we finally express $\mu$ and $\sigma^2$ as some functions  
of the discrete parameters 
$m_1$, $m_2$, $n_1$, $n_2$, and $N$:  
\beq
\mu=\left(1-\frac{\tau_3}{\tau_1^2}\right)\frac{N}{2}(N-1)
+\left(1-\frac{\tau_2}{\tau_1}\right)\frac{N}{2}, 
\label{f3}
\eeq
\begin{align}
\sigma^2 
=\frac{1}{36}
\left(1-\frac{\tau_3^2}{\tau_1^4}\right)N(4N^2+6N-1) 
+\frac{32mn}{9\tau_1^4} \,N(N-1)(N-2) \hspace{1cm}  \nn \\
 + \, \frac{(\tau_1 \tau_4-\tau_2 \tau_3)}{\tau_1^3}\,N(N-1)
+ \frac{1}{4\tau_1^2} \left(\frac{\tau_3^2}{\tau_1^2}
-\tau_2^2\right)N.    
\label{sigma1}
\end{align}
Since the Gaussian distribution (normalized to unity) corresponding
these $\mu$ and $\sigma$ is given by 
\beq
G(\mathcal{E})=\frac{1}{\sqrt{2 \pi}\sigma}
e^{-\frac{(\mathcal{E}-\mu)^2}{2\sigma^2}} \, , 
\label{f5}
\eeq
now it is possible to easily check whether
the normalized level density of the spin chain \eq{b14}
satisfies the condition $d_i \simeq G(\mc{E}_i)$ for sufficiently
large numbers of lattice sites. Indeed, 
by taking different sets of positive integer values of 
$m_1, ~m_2,~ n_1$ and $n_2$ satisfying the conditions 
$\v m_1-m_2 \v > 1 $ and $\v n_1-n_2 \v > 1 $,
we find that the normalized level density of the spin chain \eq{b14}
is in excellent agreement with the Gaussian distribution \eq{f5} 
for moderately large values of $N$ ($N\geq 15$). 
As an example, in Fig.~1 we compare the normalized level density 
with the Gaussian distribution for the case 
$m_1 = 3$, $m_2 = 1$, $n_1 = 4$, $n_2 = 1$ and $N=20$. 
We also calculate the mean square error (MSE)
for the above mentioned case and find it to be as low as $1.34 \times 10^{-8}$.
Furthermore, this MSE reduces to $1.86\times 10^{-10}$ when we take 
$N=40$ and keep all other parameters unchanged. Thus
the agreement between normalized level density 
of the spin chain \eq{b14} and the Gaussian distribution 
\eq{f5} improves with the increasing value of $N$.
\begin{figure}[htb]
\begin{center}
\resizebox{100mm}{!}{\includegraphics{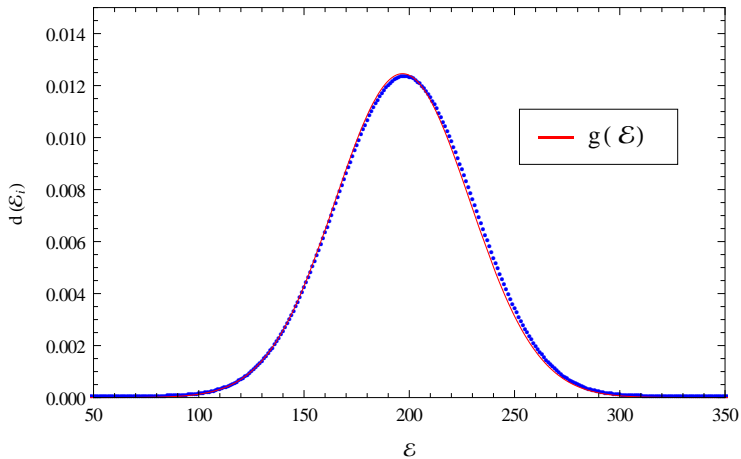}}
{\small{\caption{Continuous red curve represents 
the Gaussian distribution and blue dots represent the  
level density distribution of the spin chain \eq{b14} with 
$m_1 = 3$, $m_2 = 1$, $n_1 = 4$, $n_2 = 1$ and $N=20$.}}}
\label{f4}\end{center}
\end{figure}

Next, we shall study the 
distribution of spacing between
consecutive energy levels for the spin chain \eq{b14}. 
For the purpose of eliminating the effect of local
level density variation in the distribution of spacing between
energy levels, an unfolding mapping is usually 
employed to the `raw' spectrum \cite{Ha01}.
Since the level density of the spin chain 
\eq{b14} obeys  Gaussian distribution for large number of lattice sites, 
one can express the corresponding cumulative level density 
$\eta(\mc{E})$ through the error function as
\beq
\eta(\mathcal{E})= \int_{-\infty}^{\mathcal{E}} G(x) dx =
\frac{1}{2}\left[1+\rm{erf}\left(\frac{\mathcal{E}-\mu}
{\surd{2}\sigma}\right)\right]. 
\label{f7}
\eeq
For the case of spin chain \eq{b14}, this 
cumulative level density function is applied to map the
energy  levels $\mathcal{E}_i,~~ i = 1 , . . . , l$, into 
unfolded energy levels of the form $\eta_i \equiv \eta(\mathcal{E}_i)$.
The cumulative level spacing distribution for such unfolded energy
levels is obtained through the relation 
\beq
P(s) = \int_0^s p(x) dx \, ,
\label{f6} 
\eeq
where $p(s_i)$ denotes the probability density 
of normalized spacing $s_i$ given by 
$s_i = (\eta_{i+1} -\eta_i) / \Delta$ and   
$\Delta = (\eta_l-\eta_1) /(͑l-1)$  
is the mean spacing between unfolded energy levels. 
According to a well-known 
conjecture by Berry and Tabor, 
the density of normalized spacing  
for a `generic' quantum integrable system
should obey the Poisson's law given by $p(s) = e^{-s}$ 
\cite{BT77}. However, it has been observed 
earlier that $p(s)$ does not exhibit this Poissonian behaviour 
for a large class of quantum integrable spin
chains with long-range interactions
\cite{FG05,EFGR05,BFGR08,BFGR08epl,BB09,BBB14}. 

To explain the above mentioned anomalous behaviour in the spectra of 
quantum integrable spin chains with long range interactions,
it has been analytically shown in Ref.~\cite{BFGR08} that 
if the  discrete spectrum of a quantum system 
satisfies the following four conditions: \\
i) the energy levels are equispaced, i.e.,
$\mathcal{E}_{i+1}-\mathcal{E}_i=\delta$, 
for $i=1, 2, \ldots, l-1$,  \\
ii) the level density is approximately Gaussian, \\
iii) $\mathcal{E}_{max}-\mu, ~\mu-\mathcal{E}_{min} \gg \sigma$, \\
iv) $|\mathcal{E}_{max}+\mathcal{E}_{min}-2\mu| \ll
\mathcal{E}_{max}-\mathcal{E}_{min} \, ,$\\
then the corresponding cumulative level spacing
distribution is approximately given by 
\beq
\tilde{P}(s)\simeq 1-\frac{2}{\sqrt{\pi}s_{max}}
\sqrt{\ln \left(\frac{s_{max}} 
{s}\right)} \, ,
\label{f8}
\eeq
where 
\beq
s_{max}=\frac{\mathcal{E}_{max}-\mathcal{E}_{min}}{\sqrt{2\pi}~\sigma} \, .
\label{f9}
\eeq
Since, the spectra of many quantum integrable spin
chains with long-range interactions 
satisfy the above mentioned
four conditions with reasonable accuracy, 
the cumulative level density of such spin chains  
obey the `square root of a logarithm' law \eq{f8}. 
In the case of presently considered spin chain \eq{b14}, 
it has been already found that the conditions i) and ii) are satisfied.
For the purpose of analyzing the remaining conditions, we use  
Eqs.~\eq{e3}, \eq{e4}, \eq{e5} and \eq{e6} to obtain  
$\mathcal{E}_{min}=O(N)$ and
$\mathcal{E}_{max}=N^2 + O(N)$. Moreover,
with the help of Eqs.~(\ref{f2}) and (\ref{sigma}), 
we find that 
\beq
\mu=\frac{1}{2} \left(1-\frac{\tau_3}{\tau_1^2}\right)N^2 +
O(N), ~~~~
\sigma^2=\frac{1}{9}
\left[1-\frac{\tau_3^2-32mn}{\tau_1^4}\right]N^3+O(N^2).  \nn
\eeq
\begin{figure}[htb]
\begin{center}
\resizebox{100mm}{!}{\includegraphics{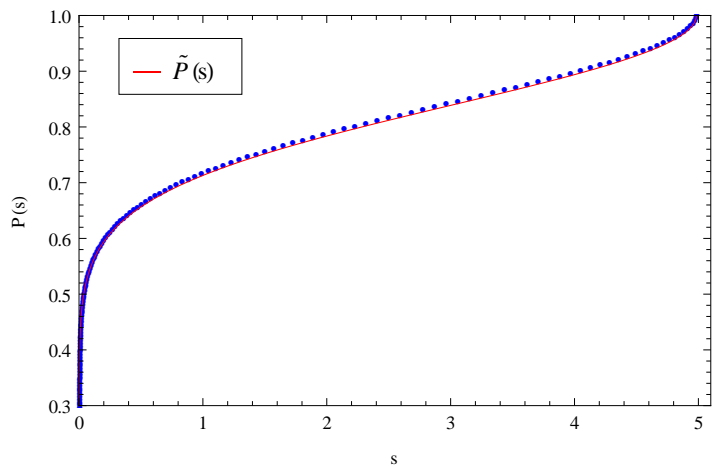}}
{\small{\caption{Blue dots represent cumulative level
spacing distribution $P(s)$ for the spin chain with
$m_1 = 3$, $m_2 = 1$, $n_1 = 4$, $n_2 = 1$ and  
$N=20$, while continuous red line is the corresponding
analytic approximation 
$\tilde{P}(s)$.
}}}
\label{test1}
\end{center}
\end{figure}
Since $\tau_1=m+n$ and $\tau_3=m-n$, the leading order 
contributions to mean and variance in the above equation
interestingly depend only on the values of $m$ and $n$. 
Using the leading order contributions to $\mathcal{E}_{min}$, 
$\mathcal{E}_{max}$, $\mu$ and $\sigma^2$, it is easy to check that 
the conditions iii) is also obeyed 
for the spectrum of the spin chain \eq{b14} with
$N\gg 1$, whereas condition iv) holds only in the case when $m=n$. 
However, it can be shown that even if condition iv) is dropped, 
Eq.~(\ref{f8}) is still
obeyed within a slightly smaller range of $s$~\cite{BFGR09}. 
Hence, it is expected that $P(s)$ in \eq{f6} would
follow the analytical expression 
$\tilde{P}(s)$ in \eq{f8} for the case of spin chain \eq{b14}. 
With the help of Mathematica, we compute $P(s)$
by taking different sets of positive integer values of 
$m_1,~ m_2,~ n_1$ and $n_2$ satisfying the conditions 
$\v m_1-m_2 \v > 1 $ and $\v n_1-n_2 \v > 1 $, and for 
moderately large values of $N$. 
It turns out that $P(s)$ obeys 
the analytical expression \eq{f8}    
with remarkable accuracy in all of these cases. 
As an example, in Fig.~2 we compare
$P(s)$ with $\tilde{P}(s)$ for the particular case   
$m_1 = 3$, $m_2 = 1$, $n_1 = 4$,  $n_2 =1$ and $N=20$. 

\bigskip

\noi \section{Conclusions}
\renewcommand{\theequation}{8.{\arabic{equation}}}
\setcounter{equation}{0}
\medskip
Here we construct   
SAPSRO which satisfy the $BC_N$ type of Weyl algebra
and lead to a novel class of spin Calogero models 
as well as related PF chains with reflecting ends. 
We compute the exact spectra of these $BC_N$ type of spin 
Calogero models, by using the fact that 
their Hamiltonians can be represented in triangular forms  
while acting on some partially ordered
sets of basis vectors of the corresponding Hilbert spaces. 
Since the strong coupling limit of these spin 
Calogero models yields $BC_N$ type of PF chains with SAPSRO, 
we apply the freezing trick to obtain the partition functions 
of this type of PF spin chains in a closed form. 
We also derive a formula \eq{d6} which expresses 
such a partition function 
in terms of known partition functions of several $A_{K}$
type of supersymmetric PF spin chains, where $K\leq N-1$. 
By using this formula,
we analyze statistical properties like 
level density distribution and nearest neighbour  
spacings distribution
in the spectra of spin chains  
with sufficiently large number of lattice sites. 
It turns out that, in analogy with the case of many other 
integrable systems with long-range interactions,
the level density of  PF spin chains with SAPSRO 
follows the Gaussian distribution 
and the cumulative nearest neighbour spacings distribution  
obeys the `square root of a logarithm' law.

In this paper, we show that the partition functions 
of PF spin chains with SAPSRO obey an interesting type of 
duality relation. 
To this end, we consider a new quantum number
which measures the parity of the spin states under the action of SAPSRO.
It is found that 
the partition functions of these spin chains
satisfy an `extended' boson-fermion duality relation \eq{g13},  
which involves not only the exchange of bosonic and 
fermionic degrees of freedom, but also the exchange of           
positive and negative parity degrees of freedom associated with 
SAPSRO. As an application of this duality relation, we  
compute the highest energy levels of these spin chains
from their ground state energies.
Moreover, we find that partition functions of a  large 
class of integrable and nonintegrable 
spin chains with Hamiltonians of the form \eq{g14} satisfy  
this type of duality relation. 

We have mentioned earlier that, $BC_N$ type of PF spin chains 
with SAPSRO do not 
exhibit global \su{m|n} supersymmetry for arbitrary 
values of the related discrete parameters. However, for a  
particular choice of these discrete parameters, 
SAPSRO reduce to the trivial 
identity operator and lead to the \su{m|n} supersymmetric
Hamiltonian $\mathcal{H}^{(m,0|n,0)}$
in \eq{b16}. Curiously, we find that the partition function and spectrum 
of this $\mathcal{H}^{(m,0|n,0)}$ coincide with those   
of $A_{N-1}$ type of \su{m|n} supersymmetric PF chain
with Hamiltonian $\wt{\mc{H}}_\mr{PF}^{(m|n)}$ in \eq{scal}.
Consequently, these two Hamiltonians are related
through a unitary transformation of the form \eq{sim} and  
the spectrum of $\mathcal{H}^{(m,0|n,0)}$
can be expressed through Haldane's motifs as given in \eq{motif}. 
As a future study,
it would be interesting to find out whether some modification 
of these motifs can be used to describe the spectra of 
$BC_N$ type of PF spin chains 
with SAPSRO for other possible choices of the related discrete
parameters. It may also be noted that, 
$A_{N-1}$ type of  PF chain with Hamiltonian 
$\wt{\mc{H}}_\mr{PF}^{(m|n)}$ in \eq{scal} exhibits the 
$Y(gl(m|n))$ super Yangian symmetry~\cite{Hi95npb,HB00}.
Hence, due to the existence of unitary transformation
\eq{sim}, it is evident that the Hamiltonian  
$\mathcal{H}^{(m,0|n,0)}$ also exhibits the 
$Y(gl(m|n))$ super Yangian symmetry. However, finding out the 
explicit form for the conserved quantities of $\mathcal{H}^{(m,0|n,0)}$, 
which would satisfy the $Y(gl(m|n))$ algebra, 
remains an interesting problem on which 
work is currently in progress.

\bigskip 
\noi {\bf Acknowledgements}
\medskip

The authors thank Artemio Gonz\'alez-L\'opez and Federico Finkel 
for fruitful discussions. One of the authors (B.B.M.) thanks the Abdus 
Salam International Centre for Theoretical Physics 
for a Senior Associateship, which partially supported this work.

\bigskip

\bibliographystyle{model1a-num-names}
\bibliography{cmprefs}

\end{document}